\documentclass[aps]{revtex4}
\usepackage{amsmath,amssymb,graphicx,mathrsfs,hyperref}

\newcommand{\be}{\begin{equation}}
\newcommand{\ee}{\end{equation}}
\newcommand{\bea}{\begin{eqnarray}}
\newcommand{\eea}{\end{eqnarray}}

\def\({\left(} \def\){\right)}
\begin{document}

\title
{Graviton multi-point functions at the
 AdS boundary}
\author{ Ram Brustein${}^{(1)}$,  A.J.M. Medved${}^{(2)}$ \\
(1)\ Department of Physics, Ben-Gurion University,
    Beer-Sheva 84105, Israel \\
(2)  Department of Physics \& Electronics, Rhodes University,
  Grahamstown 6140, South Africa 
 \\ 
    ramyb@bgu.ac.il,\  j.medved@ru.ac.za
}

\begin{abstract}
{The gauge-gravity duality can be used to relate connected multi-point graviton functions to connected multi-point correlation functions of the stress tensor of  a strongly coupled fluid. Here, we show how to construct the connected graviton functions for a particular kinematic regime that is ideal for discriminating between different gravitational theories; in particular, between Einstein theory and its leading-order string theory correction. Our analysis begins with the one-particle irreducible graviton amplitudes in an anti-de Sitter black brane background. We show how these can be used to calculate the connected graviton functions and demonstrate that the two types of amplitudes agree in some cases. It is then asserted on physical grounds that this agreement persists in all cases for both Einstein gravity and its leading-order  correction. This outcome implies that the corresponding field-theory correlation functions can be read directly off from the bulk Lagrangian, just
as can be done for the ratio of the shear viscosity to the entropy density.}
\end{abstract}
\maketitle

\section{Introduction}

The gauge--gravity duality implies that a strongly coupled gauge field theory in four dimensions should have a dual description as a weakly coupled gravity theory in  five-dimensional anti-de Sitter (5D AdS) space \cite{Maldacena1,Maldacena}. Because the gauge theory lives at the outer boundary of the AdS bulk spacetime, this is also a bulk--boundary or ``holographic'' correspondence \cite{Witten}.

String theory is a unitary and UV-complete theory. Consequently, any consistent effective description of the physics, on either of the side of the duality, must be unitary. This observation lead us to the following conclusion \cite{prepap}: An effective theory describing small perturbations about the gravitational background  solution cannot have more than two time derivatives in its linearized equation of motion. In other words,  such an  effective theory of gravity  must either be Einstein's two-derivative theory, a member of the exclusive  Lovelock class of higher-derivative theories \cite{LL} or else be supplemented by boundary conditions that  enforce this two-derivative limit. We will often use a term like ``Lovelock'' or ``Gauss--Bonnet''  (the four-derivative Lovelock theory) with the understanding  that  either of the latter two scenarios is realized. See Section~1.2 of \cite{cavepeeps} for further discussion.

Lovelock gravity consists of an infinite series of terms that are  arranged in order of  increasing numbers of pairs of derivatives. But, given a sensible perturbative expansion, it is likely that the correction to the Einstein Lagrangian
at leading order will be the most easily accessed by experiment (if at all). So that it makes sense to say that the gravitational dual is   Einstein  gravity plus a  Gauss--Bonnet correction, or possibly just Einstein gravity. We will be proceeding with this as our premise.

Our claims about the gravitational dual can be put to the test,
as we now elaborate on. In a  recent publication \cite{cavepeeps}, we systematically studied  gravitational   perturbations
about a black brane solution  in 5D AdS space.
We focused on a certain kinematic region of ``high momentum'', which will be clarified early in the paper, but otherwise
computed all of the physical, on-shell one-particle irreducible (1PI) graviton $n$-point amplitudes for both
Einstein gravity and its leading-order Gauss--Bonnet correction.
Some of the more important findings are listed
as follows:  \\
1) All 1PI $n$-point functions with $n$ odd are trivially vanishing. \\
2) All 1PI $n$-point functions involving the scalar graviton  modes (or the sound channel \cite{PSS2}) are trivially vanishing. \\
3) All 1PI $n$-point functions involving the vector graviton modes (or the shear channel \cite{PSS2})
are essentially featureless and theory independent. \\
4) Except for the two-point function, any 1PI $2n$-point function
consisting strictly of  tensor graviton  modes depends on the
scattering  angles and in a theory-dependent way. \\
5) For Einstein gravity, the (tensor) four- and higher-point
functions all go as $s$,
where $s$ has its
usual meaning as the Mandelstam center-of-mass variable.   \\
6) The contribution of the leading-order  Gauss--Bonnet correction goes as $s^2$
for the four-point function and $s(s+v)$ for higher-point functions, where  $v$
is an independent angle that can be interpreted as
a ``generalized Mandelstam'' variable.

In brief, what we have found is that  both Einstein gravity and its leading-order Gauss--Bonnet correction exhibit a universal angular dependence in multi-point graviton functions (at least for a particular kinematic region). The two extra derivatives in Gauss--Bonnet gravity  lead to a more complicated angular dependence, which would become increasingly more complex as the perturbative order of corrections is increased. But the added complexity only becomes  evident at the six-point (and higher) level  and, even then, only for a certain class of gravitons.

The black brane geometry  provides a framework for probing the hydrodynamic properties of a strongly coupled fluid \cite{KSS-hep-th/0309213,SS,hydrorev}, such as the quark--gluon plasma \cite{QCD}. The gauge--gravity duality implies that there should be an analogous list of statements to the above list about the stress-tensor correlation functions of a strongly coupled gauge theory.  This point was already mentioned in our  previous work \cite{cavepeeps} and has since been made concrete in a more recent article \cite{newby}.

An obstacle to directly applying  the duality for this purpose is that the graviton amplitudes of \cite{cavepeeps} are the 1PI amplitudes. What is rather needed to make contact with the gauge theory would be the connected graviton functions. The reason being that the 1PI functions on the field-theory side are not particularly useful because of complications that arise in separating these from the connected functions (see \cite{Showk-Pap,Pap} for a recent discussion). The best way to circumvent such issues is to have things already worked  out in the bulk before applying  the duality.

The main objective of the current study is to fill in this gap on the bulk-to-boundary chain and make clear how the connected functions can be calculated from the 1PI ones. Working  in the high-momentum kinematic region, we  are able to show  that the four-, six- and eight- point connected functions for Einstein gravity are the same as the 1PI ones. Similar agreement is established for the four- and six-point functions when the leading-order Gauss--Bonnet corrections are included.

Following our presentation of explicit calculations, we will assert, using general covariance and additional general arguments, that this agreement between the connected and 1PI functions should persist for the higher-point amplitudes as well. This is an interesting outcome in its own right but  is especially significant in the context of the gauge--gravity duality.

What our findings make clear is that, at least for the high-momentum regime, the multi-point correlation functions of the gauge-theory stress tensor fall into universality classes \cite{newby} in the same way that the ratio of the shear viscosity to the entropy density famously does \cite{PSS,KSS}. And, just as one can read off this ratio of two-point correlators directly from the bulk Lagrangian \cite{BMold,Iqlui,Cai}, the same can be done for the higher-point correlation functions as well. Importantly, this provides a new means for probing the gravitational dual of a strongly coupled gauge theory.

The rest of the paper proceeds as follows:
The next section clarifies our choice of kinematic region and Section~3 reviews some of the important results from our previous article \cite{cavepeeps}. In particular, we recall our formulations for the 1PI graviton functions and discuss
how these are related to correlation functions of the boundary theory. Section~4 is where we begin to address the relation between the connected and 1PI amplitudes, starting with a preliminary discussion about the combinatorics of Feynman diagrams. In Section~5, we show by explicit calculation that the one-particle-reducible (1PR) contribution to the connected six-point function  identically vanishes (the four-point case is trivial). This outcome is  extended to all the reducible diagrams with a single internal line in Section~6. Then, in Section~7, we present our argument as to why these cancelations should universally persist. Section~8 considers disconnected graviton functions, which can  also  be useful for probing the gravitational dual. Section~9 ends the main text with a brief overview and Appendix~A includes a calculation that is referred to in Section~7.

Let us briefly mention coordinate conventions. We assume
a 5D bulk spacetime and use
 $\;{\bf x}=\left(t,x,y,z\right)\;$
to denote  the coordinates of the four-dimensional Poincar\'{e}-invariant
subspace of the brane. Without loss of generality, $z$ is designated
as the direction of  graviton propagation along
the brane (or along any other spacetime slice
at constant radius). For indices,  $\;a,b,\cdots=\left\{t,x,y,z\right\}\;$.
The radial coordinate  is denoted by $r$ and
ranges  from $\;r=r_h\;$ at the black brane horizon
to $\;r\to\infty\;$ at the AdS boundary.

\section{The high-momentum regime}

Before continuing, we wish to clarify how the
kinematic region of high momentum is defined. Here, the focus is on the bulk perspective; for the boundary point of view, see \cite{newby}.

Essentially, the frequencies and momenta ($\omega$'s and $k$'s) of the gravitons should be large enough so that these dominate over { all} radial derivatives, whether acting on the gravitons { or} the background. In practice, this amounts to requiring that { any} derivative acts on a graviton to produce an $\;\omega=-i\nabla_t=-i\partial_t\;$ or a
$\;k=-i\nabla_z=-i\partial_z\;$.~\footnote{The exception to this rule
is when the number of derivatives exceeds the number of gravitons,
as such a contribution would  necessarily vanish  on-shell.
Any such excess derivatives are
regarded as radial derivatives acting on the background.}
This kinematic regime is most suitable for distinguishing between Einstein gravity and higher-derivative corrections, as it emphasizes the number of derivatives in an interaction vertex.

The parameter  $\epsilon$ will serve as  a dimensionless perturbative
coefficient which measures the strength of the higher-derivative corrections to the Einstein Lagrangian, such that the Gauss--Bonnet  correction is linear in $\epsilon$.~\footnote{It is assumed that $\epsilon$ is fixed precisely  by reading off the numerical coefficient of the Riemann-tensor-squared term in the Lagrangian. The value of $\epsilon$ can also be determined through its  connection with  the shear viscosity to entropy density ratio, $\;\frac{\eta}{s}=\frac{1}{4\pi}\left[1-8\epsilon\right]\;$
in natural units \cite{BLMSY-0712.0805,same_day}.} Then  the number of derivatives is given by  $2^{j+1}$ for a term of order $\epsilon^j$. The restriction  to high momenta is really the same as asking that the Mandelstam invariant $s$ be larger than the AdS bulk curvature or $\;s\gg 1/L^2\;$, where $L$ is the AdS curvature length.

As for an upper bound, the momenta  should still be small enough to ensure that any higher-order Lovelock terms  make parametrically small contributions to  physical quantities. This is necessary  for keeping the perturbative expansion under control and, more importantly, for consistency with treating string theory as an effective theory of gravity. Now consider that each additional order in the Lovelock expansion brings with it another factor of $\epsilon$. For instance, a term  with six derivatives can contribute one more factor of $s$ but at the cost of one more factor of $\epsilon$. Hence, we require $\;s\epsilon\ll 1/L^2\;$. And, because it follows on general grounds that $\;\epsilon\sim l_p^2/L^2\ll 1\;$ ($l_p$ is the 5D Planck length and the right-hand side is a necessary requirement for the viability of the gauge--gravity duality), the previous inequality means $\;s\ll 1/l_p^2\;$. That is, all momenta must be ``sub-Planckian''.

This last constraint is rather trivial, but there is still
another one to consider. Because a black brane is dual to  a fluid, we are working
(at least implicitly) in the
hydrodynamic regime, and so the bound $\;s\ll (\pi T)^2\;$ is also required. Here, $T$ is the fluid temperature,
which is equivalent to the Hawking temperature on the brane.
We then end up with the high-momentum region being defined
by the finite range
\be
 1 \;\ll\; L \sqrt{s} \;\ll\; \pi T L \;.
\label{bounds}
\ee

There is no danger of  the right-hand inequality being in conflict with the
bulk solution because the gauge--gravity duality can only be applied consistently
in a black hole (or brane) background when the horizon is much bigger than the AdS
scale $L$ or, equivalently,
much
 hotter
than the surrounding AdS space \cite{Witten2}.
Since $\;r_h\sim \pi T L^2\;$, it follows that $\; \pi T L\gg 1\;$.

An additional comment is in order: Let us recall the relation $\lambda=g^2_{YM}N\;$ between gauge-theory parameters, where $\lambda$ is the 't Hooft coupling, $g_{YM}$ is the Yang--Mills coupling and $N$ is the number of colors.  The holographic dictionary applies when $\;N\to\infty\;$ and  $\;g_{YM}\to 0\;$,
with $\lambda$ large but finite. Hence, we  carry out our analysis at tree-level
(loop diagrams are suppressed by inverse powers of $N$) and
the leading-order modifications to Einstein gravity must
then be at higher orders in
$\lambda^{-1}$. This is consistent with looking at higher-derivative
extensions to the effective string-theory Lagrangian \cite{robtalk}.

\section{Graviton multi-point functions}

In this section, we recall  some relevant parts of our earlier work \cite{cavepeeps} as well as set up the basic framework.

The initial task is to calculate all the physical, 1PI on-shell graviton multi-point amplitudes for an  (asymptotically)  AdS theory of gravity. We assume that the background solution is a 5D black brane with a Poincar\'e-invariant horizon because, as already mentioned, this setting is useful for learning about the fluid dynamics of the gauge-theory dual.

As also mentioned, the effective theory describing small fluctuations about the background is constrained to have  a two-derivative equation of motion. It is assumed  that this effective theory  limits to Einstein gravity  in the IR and has a sensible perturbative expansion in terms of the number of pairs of derivatives.  It follows that each term in the Lagrangian is suppressed by a  factor of $\;\epsilon$ relative to its predecessor with two derivatives fewer. Hence, to leading order, we need only consider the four-derivative correction to Einstein's theory; so that the theories of interest are  Einstein gravity and  Einstein gravity plus a Gauss--Bonnet extension.

The premise is to expand the Einstein and Gauss--Bonnet Lagrangians in  numbers of gravitons. A graviton  $h_{\mu\nu}$ represents a small perturbation of the metric from its background value, $\;g_{\mu\nu}\to g_{\mu\nu} +h_{\mu\nu}\;$. The $n^{\rm th}$-order term of the  expansion can be identified with  the graviton $n$-point function, which can be represented by a 1PI  Feynman diagram. This becomes a Witten diagram \cite{Witten} in the limit that $\;r\to\;\infty\;$.

Let us introduce the following ansatz and labeling convention for the gravitons: $\;h_{\mu\nu}^{(j)} = \phi(r)\exp\left[i\omega_jt-k_jz\right]\;$. We work in the radial gauge, for which $\;h_{rr}=h_{ar}=0\;$ for any choice of $a$. This choice leads to a decoupling of the gravitons into three distinct classes \cite{PSS2}: tensors, vectors and scalars or (respectively) $\;h_2=\{h_{xy}\}\;$, $\;h_1=\{h_{zx},h_{tx}\}\;$ and $\; h_0=\{h_{tt},h_{zz},h_{zt},h^{x}_{\ x}+h^{y}_{\ y}\}\;$, with redundant polarizations omitted.

As previously motivated, we  work in the kinematic region of high
momentum, meaning that only terms with the highest allowed power of $\omega$ and $k$ are included.

Given that the calculations are on-shell and in the region of high momentum, we are able to discard any $n$-point function involving scalar modes. This, in turn, means that graviton functions with odd values of $n$ vanish, as general covariance dictates that   vectors and tensors  must come in pairs.

The case against  vectors modes is somewhat more subtle. A vector mode is analogous to a field-strength tensor \cite{KSS-hep-th/0309213}, so it must be differentiated to be physical. Moreover, any non-vanishing $n$-point function that includes vector modes is limited to exactly two derivatives \cite{cavepeeps}, both of which are necessarily constrained to act on the vectors. Because of this simplicity, this class of functions is of  limited value  to  the  current analysis and will subsequently  be ignored.

Thus, we are left with only the tensor modes.
Later on, we will calculate  connected functions for the tensors  by convolving pairs of 1PI amplitudes, and so it is important to understand why vector and scalar modes can be ignored in the  internal lines. Generally speaking, all types of  modes can appear because, unlike those appearing on external lines, internal
gravitons are not constrained to be on-shell. However, on-shell or off, the scalar modes are not physical ({\it i.e.}, can be gauged away) in the absence of an external source. (See, {\it e.g.}, \cite{spring}.) And, since  the vector modes can be identified with  eletromagnetic gauge fields \cite{KSS-hep-th/0309213},
they must likewise  be sourced to be physical.  One might then ask how our results would differ when such sources are present, as could well be the case in a string-theory context. We will make an argument that, in the high-momentum regime
and  at least to  order $\epsilon$, all internal lines -- irrespective of the internal propagating modes -- do not contribute.
We will return to this matter at the end of Section~7.

As explained in \cite{cavepeeps}, we have found the complete list of 1PI  multi-point functions for
the tensor modes at arbitrary $r$.
But, for current purposes, it is most useful to focus
on the limit of the AdS outer boundary; this
being the location of the gauge-theory dual. In this case,
\be
\lim_{r\to\infty}\langle h_2h_2 \rangle_E\;=\; \left(\frac{L}{r}
\right)^{3} k^2\;\;\;\;({\rm with}\; k^2\equiv k_1^2=k_2^2)\;,
\label{E2}
\ee
\be
\lim_{r\to\infty}\langle h_2 h_2h_2h_2\rangle_E\;=\;
\frac{9}{4}\left(\frac{L}{r}
\right)^{7}
s\;,
\label{E4}
\ee
\be
\lim_{r\to\infty}\langle h_2^{6}\rangle_E\;=\;
\frac{75}{8}\left(\frac{L}{r}
\right)^{11} s\; ,
\label{E6}
\ee
\be
\lim_{r\to\infty}\langle h_2h_2 \rangle_{GB}\;=\; \left(\frac{L}{r}
\right)^{3}\left[1-8
\epsilon\frac{L^2}{r^2}\right]
 k^2\;\;\;\;({\rm with}\; k^2\equiv k_1^2=k_2^2)\;,
\label{GB2}
\ee
\be
\lim_{r\to\infty}\langle h_2 h_2h_2h_2\rangle_{GB}\;=\;
\frac{3}{4}\left(\frac{L}{r}
\right)^{7}
\;\left[3s+\epsilon\frac{L^2}{r^2} s^2\right]\; ,
\label{GB4}
\ee
\be
\lim_{r\to\infty}\langle h_2^6 \rangle_{GB}\;=\;
\frac{15}{8}\left(\frac{L}{r}
\right)^{11}\;\left[5s+ 21 \epsilon\frac{L^2}{r^2} s(s+v)\right]\; ,
\label{GB6}
\ee
where $E$ and $GB$ respectively denote Einstein and Gauss--Bonnet corrected gravity,  the 5D Newton's constant has been fixed according to $\;16\pi G_5=1\;$, and the kinematic variables $s$ and $v$ are made explicit below. We have also employed the boundary limit of the AdS brane metric,
$\;\lim\limits_{r\to\infty} ds^2 = -\frac{r^2}{L^2}dt^2 +\frac{L^2}{r^2}dr^2 +\frac{r^2}{L^2}\left[
dx^2 +dy^2 + dz^2\right]\;$.

More generally,
for $\;2n=4,6,8\dots\;$,
\be
\lim_{r\to\infty}\langle h_2^{2n}\rangle_E \;=\;
\frac{(2n-1) \Gamma\left[n+\frac{1}{2}\right]}{\sqrt{\pi}
\Gamma[n-1]}\left(\frac{L}{r}
\right)^{4n-1} s\;,
\ee
\be
\lim_{r\to\infty}\langle h_2^{2n} \rangle_{GB} =
\lim_{r\to\infty} \langle h_2^{2n} \rangle_{E}
+
\frac{2}{5}\epsilon\dbinom{2n}{4} \frac{\Gamma\left[n+\frac{3}{2}\right]}{\sqrt{\pi}
\Gamma[n-1]}\left(\frac{L}{r}
\right)^{4n+1} s(s+v)\;,
\label{GGen}
\ee
with the understanding that $\;v=0\;$ for $\;2n=4\;$.

The external gravitons are symmetrized in our expressions, with
the center-of-mass variable $s$ and the generalized Mandelstam variable  $v$ accounting for the symmetrization:
\be
s\;=\;-\frac{1}{2n(2n-1)}\sum_{i=1}^{2n}\sum_{\overset{j=1}{j\neq i}}^{2n} k_i^{\mu}k_{j\mu}\;,
\label{mandel}
\ee
\be
v\;=\;-\frac{1}{2n(2n-1)(2n-2)(2n-3)}\sum_{i_1=1}^{2n}\sum_{\overset{i_2=1}{i_2\neq i_1}}^{2n}
\sum_{\overset{i_3=1}{i_3\neq i_{1,2}}}^{2n}\sum_{\overset{i_4=1}{i_4\neq i_{1,2,3}}}^{2n}
\sum_{\overset{j=1}{j\neq i_{1,2,3,4}}}^{2n}
 k_{i_1}^{\mu}k_{j\mu}\;.
\ee

The Gauss--Bonnet corrections depend only on the Riemann-tensor-squared term, as the other four-derivative
terms (Ricci-tensor-squared and Ricci-scalar-squared)  can be transformed away with a suitable choice of field redefinitions \cite{tH,POL,cavepeeps}. Also, to arrive at the above expressions, we  employed the transverse/traceless
gauge for the tensors (which is  compatible with our previous choice of radial gauge).

Let us briefly explain the above outcomes from a Feynman-diagram perspective: With knowledge that these are really the 1PI  graviton amplitudes, we are essentially computing a functional derivative of the form (see, {\it e.g.}, \cite{Burgess})
\be
\langle h^{2n}\rangle \;=\;(-i)^{2n+1}\frac{\delta^{2n}}{(\delta J)^{2n}}\ln\Bigg[\int {\cal D}[h]\;
 e^{i\sqrt{-g}{\cal L}[g,\nabla,h]+iJh}\Bigg]
\;,
 \ee
where an $h$ should  be regarded as a tensor, $J$ is an external source, there is an implied spacetime integral in the exponent and, since  tensor modes  come in pairs, the Lagrangian density in the exponent can be expanded out as
\be
\sqrt{-g}{\cal L}[g,\nabla,h]= {\cal L}_0[g] +\frac{h^2}{2!}{\cal L}_2[g,\nabla] +
\frac{h^4}{4!}{\cal L}_4[g,\nabla]
+\frac{h^6}{6!}{\cal L}_6[g,\nabla]  +\dots\;.
\ee

Now, taking the variations and then setting $J=0$, we have
\be
\langle h^{2n}\rangle \;=\; -i\frac{\int {\cal D}[h]\; h^{2n}
 e^{i\sqrt{-g}{\cal L}[g,\nabla,h]}}
{\int {\cal D}[h] \; e^{i\sqrt{-g}{\cal L}[g,\nabla,h]}}\;.
 \ee
 A Taylor expansion  of the exponentials then leads to the outcome
\be
\langle h^{2n}\rangle \;=\; \frac{1}{(2n)!}
\frac{\int {\cal D}[h]\; h^{2n}\left(h^{2n}{\cal L}_{2n}[g,\nabla]\right)
 e^{\frac{i}{2}h^2{\cal L}_2[g,\nabla]}}
{\int {\cal D}[h]\; e^{\frac{i}{2}h^2{\cal L}_2[g,\nabla]}}\;+\;\ldots\;,
\label{expansion1}
 \ee
with the ellipsis indicating contributions that are not
1PI.~\footnote{Alternatively, one can use the appropriate
Legendre transformation to isolate the 1PI contributions.}

To proceed, one contracts together the pair of
monomials (each of order $2n$) within the upper integral
of Eq.~(\ref{expansion1}). This can be done in $(2n)!$ different ways,~\footnote{This is because the gravitons
are contracted in pairs; one from each monomial.  The combinatorics of such contractions is discussed in Section~4.}
with this number canceling the graviton symmetrization factor in the denominator.  After the standard process of amputating the external lines,
the result of the contraction is simply ${\cal L}_{2n}(g,\nabla)$.
Equations~(\ref{E2})-(\ref{GGen}) are these quantities in momentum space.

An important feature of these multi-point functions is how they depend on the scattering angles of the gravitons. As evident from the above expressions, the two-point function has no angular dependence (an outcome that follows from momentum conservation), whereas the higher-point functions do depend on the angles, with distinct signatures for the Einstein and Gauss--Bonnet corrected parts.

This distinction  between the Einstein and Gauss--Bonnet angular dependence can be attributed to the two extra derivatives available in the Gauss--Bonnet part of the Lagrangian. The point is that both Einstein and Gauss--Bonnet gravity have a two-derivative (linearized) field equation but only Einstein gravity is truly a two-derivative theory. This is what restricts and simplifies the higher-point functions of Einstein gravity \cite{Mald-Hof,Hof}. On the other hand, the  four- and higher-point functions of any Lovelock theory will generally be  sensitive to the additional derivatives and, therefore, not subject to the same degree of constraint. Still, one can constrain any Lovelock theory to some degree by limiting the order of $\epsilon$, just like we are doing here.

\subsection{The gauge-theory perspective}

We are interested  in the stress-tensor multi-point correlation functions  for the boundary gauge theory \cite{newby}. The gauge--gravity duality implies that the stress-tensor correlation functions are  related to the connected  graviton $n$-point functions. To make the connection explicit, an  appropriate process of holographic renormalization has to be applied. This process is straightforward in our case because, in the high-momentum regime, the connected and 1PI graviton amplitudes agree. This will be shown explicitly for several cases, after which a general argument will be provided.

To put this on a more formal level, we need to apply the standard rules of holographic renormalization \cite{dBVV,skenderis1,skenderis2} to the bulk multi-point functions. This is essentially a three-step procedure: The first step (which has  already been carried out) is to  extrapolate the bulk quantity to the boundary.~\footnote{Formally, this
step requires evaluating at some large but finite value of radius
$\;r=r_{0}\;$ and then imposing the limit $\;r_0\to\infty\;$
only at the end.} The second step
is to  multiply the extrapolated functions by  a factor $\Omega^q$, where $\Omega$ is an appropriate conformal factor and
the power $q$ is determined by the conformal dimension of the operators whose correlation function is being calculated. The  third step is to subtract off any divergences
by using suitable boundary counter-terms (these essentially compensate for contributions from the background geometry).
One then only retains the part that survives in the $r\to \infty$ limit.

Let us elaborate on the second step. The conformal factor can be deduced from the asymptotic form of the AdS metric (see below Eq.~(\ref{GB6})). From this, the appropriate conformal factor can now be identified as $\;\Omega=r/L\;$, and one then  multiplies by $\Omega^q$ such that $\;q=\Delta-3\;$. Here, each  operator of (mass) conformal dimension $\Delta_i$ contributes this amount to $\Delta=\sum\limits_{i}\Delta_i$, while the subtraction of 3 is meant to ``strip off'' the metric determinant. The operators in our case are the gravitons for which  $\;\Delta_h=2\;$ (this can be deduced from the boundary behavior of the metric)  and the derivatives for which $\;\Delta_{\nabla}=1\;$.

Following the described procedure, we find that all powers of $r$
and $L$ are stripped away from our previous expressions. For instance,
\bea
\langle  h_2^{6}\rangle_{E,{\bf REN}} & = &
\lim_{r\to\infty}  \Omega^{6\cdot2+2-3} \langle h_2^{6}\rangle_E \nonumber \\
& = &
\left(\frac{r}{L}\right)^{11}\frac{75}{8}\left(\frac{L}{r}
\right)^{11} s \nonumber \\
& = &
\frac{75}{8}\;  s
\eea
and similarly for the others.

What is left is a quantity that is finite and well defined at the boundary. Hence, the first and third steps turn out to be
of no direct consequence in the specific case that we consider.
Subleading contributions to the metric could be important in principle but, in our case, only show up in terms that are asymptotically vanishing.

One might still wonder about hidden implications from the third step, as the subtraction process is the only  subtle element of the bulk-boundary dictionary. However, there are none in the high-momentum regime. To understand why, recall that the subtraction is tantamount to a process of   matching and stripping off the (divergent) bulk and boundary conformal factors, and then  eliminating contributions from the AdS background geometry. That this is the underlying premise is made clear in some of the precursory works to holographic renormalization
\cite{Witten,skenderis0,stuff,things}. And such a process would not have any bearing on our basic forms because the metric components $g^{tt}$ and $g^{zz}$ are dispersed democratically in all our amplitudes  and exhibit the same radial dependence at the boundary. Alternatively,  from the counter-term perspective, we will see later that contributions from these have no opportunity to enter into the formal calculations of the graviton amplitudes. All of this simplicity hinges heavily upon the high-momentum regime being in effect.

\section{A reminder on combinatorics of Feynman diagrams \label{combo}}

We include this discussion for completeness.
The reader who is well versed in Feynman diagrams is advised to
skip ahead to  Eq.~(\ref{connect}).

Let us start by considering a generic  $n$-point function for a general theory. Working out the Feynman combinatorics amounts to {\it (i)} collecting  the various terms that contribute to a Gaussian integral of the schematic form
\be
\int d\zeta\; \zeta^n e^{-\frac{1}{2}\zeta^2 + \frac{a}{3!}\zeta^3 +
\frac{b}{4!} \zeta^4 + \frac{c}{5!} \zeta^5 + \frac{d}{6!} \zeta^6+\cdots}\;
\ee
and  then {\it (ii)} enumerating the different ways of pairing
 up the  $\zeta$'s.

The monomial $\zeta^n$ represents the external legs and the exponent is an expansion of the Lagrangian in powers of $\zeta$. The expansion coefficients  $\;a,b,c,\dots\;$ are theory-dependent numbers that play no role in the combinatoric part of the analysis. The symmetrization factors are, however, important.

Expanding out the exponent, we have
\be
e^{-\frac{1}{2}\zeta^2}\zeta^n\left[1+\frac{a}{3!}\zeta^3+\frac{1}{2!}
\frac{a^2}{(3!)^2}\zeta^3\zeta^3+\dots\right]\times
\left[1+\frac{b}{4!}\zeta^4+\frac{1}{2!}\frac{b^2}{(4!)^2}\zeta^4\zeta^4+\dots\right]\times
\cdots \;,
\ee
with the once-exponentiated monomials  now representing interaction vertices. A term in this expansion makes a contribution  to a  full $n$-point function $\langle \zeta^n\rangle_{Ful}$  whenever it contains  an even number of $\zeta$'s, and the weight of this contribution is determined by counting  the number of distinct ways of contracting the available $\zeta$'s in pairs. In short, one is counting the associated number of Feynman diagrams.

Connected and 1PI functions are more constrained than the full functions. A connected function  $\langle \zeta^n\rangle_{Con}$ requires that two $\zeta$'s from the same monomial cannot be contracted together, whereas a 1PI function $\langle \zeta^n\rangle_{1PI}$ (or a connected function with no internal lines) further requires that any $\zeta$ from a once-exponentiated monomial
can only be contracted with a $\zeta$ from  the external monomial $\zeta^n$.

Since we are working at tree-level, even more constraints apply
to the connected functions. First consider an arbitrary term
in the expansion, which is a  product of  $m+1$ monomials,
\be
{\cal O}_{n+\sum\limits_{i=1}^m j_i}=\zeta^n \frac{\zeta^{j_1}}{j_1!} \frac{\zeta^{j_2}}{j_2!} \cdots \frac{\zeta^{j_m}}{j_m!}\;.
\ee
One can use the  ``conservation of ends'', $\;2I+E=V\;$  \cite{Burgess}, to calculate the number of internal lines $I$, given the knowledge of the number of external lines $E$ and the total number of lines ending on a vertex $V$. Here, $\;E=n\;$ and $\;V=\sum_{i=1}^{m} j_i\;$. Some simple considerations then lead to the number of loops  going as $\;L=I+1-m\;$ \cite{Burgess}. Putting this all together, one finds
\bea
L &=& \frac{1}{2}\left[\sum_{i=1}^{m}j_i \;-\;2(m-1)-n\right]\;
\nonumber \\
 &=& \frac{1}{2}\Bigg[\left(m_3+ 2m_4 + 3m_5 + 4m_6 +\dots\right)-(n-2)\Bigg]\;,
\eea
where $m_i$ counts the number of occurrences of $\zeta^i$. The tree-level constraint  means  restricting to $\;L=0\;$.

\begin{figure}[t]
\vspace{-1.5in}\hspace{.2in}
\scalebox{.55}{\includegraphics[angle=-90]{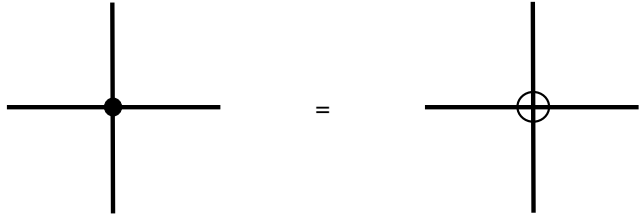}}
\caption{Four-point function. The connected function on the left is equal to the 1PI function on the right.}
\end{figure}

Let us apply this general methodology to the current scenario, starting with the graviton four-point function in the high-momentum region. Since the three-point function is identically vanishing, the only tree-level contribution in this kinematic region comes from $\;m_4=1\;$  and   $\;m_i=0\;$ otherwise. Then  $\;n=V=4\;$ and so $\;I=\frac{1}{2}\left(4-4\right)=0\;$, which identifies this as a 1PI function. Hence, it follows that
\be
\langle hhhh \rangle_{Con}\;=\;\langle hhhh
\rangle_{1PI}
\ee
must be  trivially  true. This is depicted in Fig.~1.

The first non-trivial case is the six-point function.
The tree-level possibilities are { either}
$\;m_6=1\;$ { or} $\;m_4=2\;$  and, otherwise, $\;m_i=0\;$.
The former is the 1PI six-point diagram and
the latter represents the convolution of
a pair of four-point functions, making it
a 1PR contribution to the connected function.
 These identifications lead us to
\be
\langle hhhhhh \rangle_{Con}\;=\;\langle hhhhhh
\rangle_{1PI}
+c_6\left[\frac{\langle hhhh \rangle_{Con}
\langle hhhh\rangle_{Con}}
{\langle hh\rangle}\right]\;,
\label{connect}
\ee
as shown in Fig.~2. Here,
the ${\langle hh\rangle}$ in the denominator represents
an inverse propagator  which ``compensates''
for the propagator from the internal line and
$c_6$ is a numerical coefficient that measures the relative
weight of the 1PR  contribution.
The next  step is to determine $c_6$.

\begin{figure}[t]
\vspace{-1.5in}\hspace{-.5in}
\scalebox{.55}{\includegraphics[angle=-90]{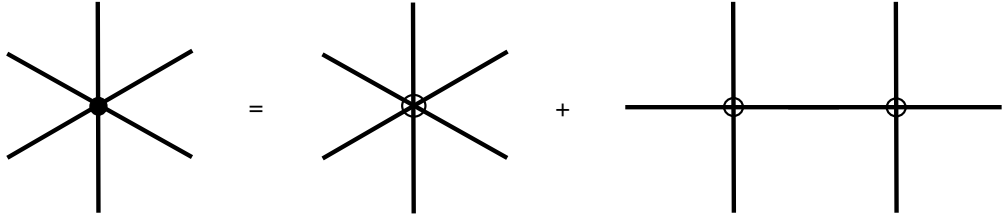}}
\vspace{-.5in}
\caption{Six-point function. The connected function has a 1PI part and a reducible part.}
\end{figure}

Applying the previously discussed rules, we find that $\;c_6=10\;$. The 1PI case requires contracting $\zeta^6 \zeta^6/6!$  with no pairs from the same monomial. This can be done in $6!$ ways, leading to $\;6!/6!=1\;$. Meanwhile, the 1PR case entails the contraction of $\zeta^6\zeta^4\zeta^4/2!(4!)^2\;$. Drawing a single $\zeta$ out of each of the $\zeta^4$'s can be done in $4^2$ ways. The drawn pair is  contracted. Then the remainder $\zeta^3 \zeta^6 \zeta^3/2!(3!)^2\;$, when subjected to the discussed rules, amounts to  $\zeta^6\zeta^6/2!(3!)^2 \;$ and so $\;6!/2!(3!)^2=10\;$ follows.

If we were evaluating the Feynman diagrams for a theory of interacting scalars, this would be the end of the calculation. However, our actual interest is   the Witten diagrams \cite{Witten} for a theory of spin-2 gravitons, and so
things are more involved.

The full calculation necessitates two additional steps. In the first, or  ``Step~1'', we momentarily  ignore all tensor indices and use holographic techniques to convolve a pair of four-point functions into an explicit six-point form. ``Step~2'' then accounts for the previously neglected tensor structure. We will discuss each of these in turn, beginning with the case of pure Einstein gravity and then the leading-order Gauss--Bonnet correction in the sequel.

\section{The 1PR six-point function...}

\subsection{...for Einstein gravity \label{einsix}}

\subsubsection{Step 1 \label{stepone}}

Let us start here with the position-space representation
of the connected or, equivalently, 1PI four-point function. This function is depicted on the right hand side of Fig.~3.
When integrated, this leads to the amplitude
\be
\langle 4 \rangle_{1PI}\;=\;\lim_{r \to \infty}\int dr \int d^4 x
\sqrt{-g(r,{\bf x})} \;h(r,{\bf x})\;h(r,{\bf x})\;\nabla_e h(r,{\bf x})
\;\nabla^e h(r,{\bf x})\;,
\ee
where all tensor structure has been suppressed (as this will be accounted for in Step~2) and the holographic limit of $\;r\to\infty\;$ has been imposed. We also suppress labeling indices on the gravitons, with the understanding that all gravitons in a 1PI function and all external gravitons in a 1PR function are symmetrized. Take notice of  the two derivatives. This number is fixed for Einstein gravity in the high-momentum regime.

\begin{figure}[t]
\vspace{-1.5in}
\scalebox{.65}{\includegraphics[angle=-90]{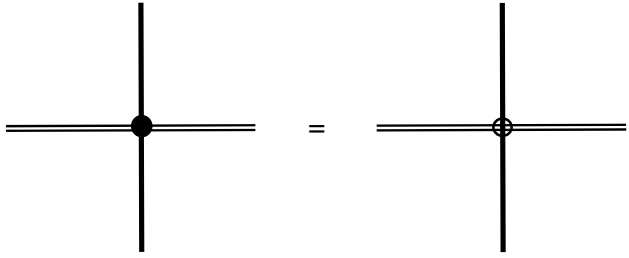}}
\vspace{-.5in}
\caption{Connected four-point function with two derivatives. The connected function is equal to the 1PI function. The differentiated gravitons are depicted as a double line and the undifferentiated gravitons, as a single line.}
\end{figure}

The current objective is to convolve a pair of these four-point functions and then see what it takes to manipulate this into an explicit six-point form. Now consider that each four-point function must ``donate'' precisely  one internal graviton; otherwise, the six-point function would not be connected. Since each four-point function carries two derivatives, there are three distinct cases: (i) the two internal gravitons are both differentiated, (ii) both undifferentiated or (iii) ``mixed'' (one is differentiated and one is not). The six-point function and the different cases are shown in Fig.~4.

\begin{figure}[t]
\scalebox{.55}{\includegraphics[angle=-90]{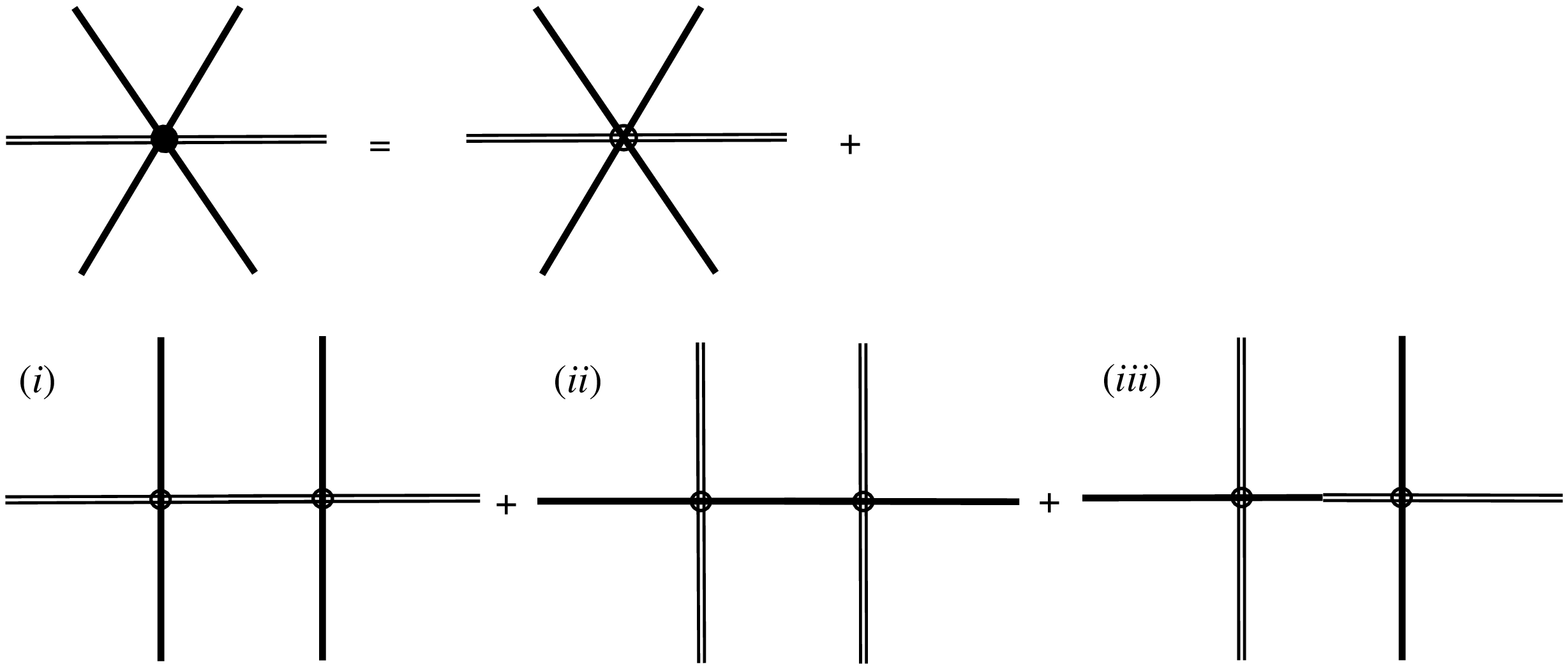}}
\caption{Six-point function. Single lines denote undifferentiated gravitons and double lines, differentiated gravitons. The three diagrams $(i)$, $(ii)$, $(iii)$ differ in the way that the internal gravitons are contracted.}
\end{figure}

Starting with the first of the three cases, we have
\be
\langle 6 \rangle^{(i)}\;=\;\lim_{r,\tilde{r} \to \infty}\int dr \int d^4x \int d\tilde{r}\int d^4\tilde{x}
\sqrt{-g(r,{\bf x})}\sqrt{-g(\tilde{r},\tilde{{\bf x}})}\;h(r,{\bf x})\;h(r,{\bf x})
\nonumber \ee \be
\times\;  \nabla_e h(r,{\bf x})
 \;\Big\langle
\nabla^e h(r,{\bf x})\;\widetilde{\nabla^{f}} h(\tilde{r},\tilde{{\bf x}}) \Big\rangle\;
\widetilde{\nabla_{f}}h(\tilde{r},\tilde{{\bf x}})
 \;h(\tilde{r},\tilde{{\bf x}})\;h(\tilde{r},\tilde{{\bf x}}) \;,
\ee
where the expectation value is used to represent the pair of internal gravitons
and a superscript
is included  on the left-hand side to distinguish between
the three  cases.

For the purpose of recasting this into a six-point form,
we first apply integration by parts to obtain
\be
\langle 6 \rangle^{(i)}\;=\;
-\lim_{r,\tilde{r} \to \infty}\int dr \int d^4x \int d\tilde{r}\int d^4\tilde{x}
\sqrt{-g(r,{\bf x})}\sqrt{-g(\tilde{r},\tilde{{\bf x}})}\;h(r,{\bf x})
\nonumber \ee \be
\times\; h(r,{\bf x})\; \nabla_e h(r,{\bf x})
 \;\Big\langle
\nabla^e h(r,{\bf x}) \;h(\tilde{r},\tilde{{\bf x}}) \Big\rangle\;
\widetilde{\nabla^{f}}\Big[\widetilde{\nabla_{f}}h(\tilde{r},\tilde{{\bf x}})
 \;h(\tilde{r},\tilde{{\bf x}})\;h(\tilde{r},\tilde{{\bf x}})\Big]\;
\ee
or, using the linearized field equation
$\;\nabla_a\nabla^a h(r,{\bf x})=0\;$~\footnote{Here, $h$ should be regarded as mixed-index graviton $h^a_{\ b}$. It behaves like a massless scalar upon covariant differentiation.} along
with the product rule,
\be
\langle 6 \rangle^{(i)}\;=\;
-2\lim_{r,\tilde{r} \to \infty}\int dr \int d^4x \int d\tilde{r}\int d^4 \tilde{x}
\sqrt{-g(r,{\bf x})}\sqrt{-g(\tilde{r},\tilde{{\bf x}})}\;h(r,{\bf x})
\nonumber \ee
\be
\times\; h(r,{\bf x})\;  \nabla_e h(r,{\bf x})
 \;\Big\langle
\nabla^e h(r,{\bf x}) \;h(\tilde{r},\tilde{{\bf x}}) \Big\rangle\;
\widetilde{\nabla_{f}}h(\tilde{r},\tilde{{\bf x}})
\;\widetilde{\nabla^{f}}h(\tilde{r},\tilde{{\bf x}}) \;h(\tilde{r},\tilde{{\bf x}})\;.
\ee

Next, we interchange the gravitons (because they are symmetric),
use  the inverse of the product rule to write
\be
\langle 6 \rangle^{(i)}\;=\;
-\frac{2}{3}\lim_{r,\tilde{r} \to \infty}\int dr \int d^4x \int d\tilde{r}\int d^4 \tilde{x}
\sqrt{-g(r,{\bf x})}\sqrt{-g(\tilde{r},\tilde{{\bf x}})}
\nonumber \ee
\be \times \;
\nabla_e\Big[h(r,{\bf x})\;h(r,{\bf x})\;h(r,{\bf x})
\Big]
\; \Big\langle
\nabla^e h(r,{\bf x}) \;h(\tilde{r},\tilde{{\bf x}}) \Big\rangle\;
\widetilde{\nabla_{f}}h(\tilde{r},\tilde{{\bf x}})
\;\widetilde{\nabla^{f}}h(\tilde{r},\tilde{{\bf x}}) \;h(\tilde{r},\tilde{{\bf x}})\;
\ee
and then integrate by parts a second time, giving
\be
\langle 6 \rangle^{(i)}\;=\;
\frac{2}{3}\lim_{r,\tilde{r} \to \infty}\int dr \int d^4x \int d\tilde{r}\int d^4 \tilde{x}
\sqrt{-g(r,{\bf x})}\sqrt{-g(\tilde{r},\tilde{{\bf x}})}\;h(r,{\bf x})
\nonumber \ee \be
\times\; h(r,{\bf x})\; h(r,{\bf x}) \;
\nabla_e\nabla^e \Big\langle h(r,{\bf x}) \;h(\tilde{r},\tilde{{\bf x}}) \Big\rangle\;
\widetilde{\nabla_{f}}h(\tilde{r},\tilde{{\bf x}})
\;\widetilde{\nabla^{f}}h(\tilde{r},\tilde{{\bf x}}) \;h(\tilde{r},\tilde{{\bf x}})\;.
\ee

Finally, we
apply the Green's function form of the field equation,~\footnote{The internal gravitons cannot be restricted
to any particular kinematic regime; meaning that, for these modes, $r$ derivatives matter. This assures the validity of Eq.~(\ref{green}).}
\bea
\nabla_a \nabla^a \Big\langle h(r,{\bf x}) \;h(\tilde{r},\tilde{{\bf x}})\Big\rangle
&=& \frac{1}{\sqrt{-g(r,{\bf x})}} \delta^{(4)}({\bf{x}}-\tilde{{\bf x}})
\delta(r-\tilde{r})\nonumber \\
&=&
\frac{1}{\sqrt{-g(\tilde{r},\tilde{{\bf x}})}} \delta^{(4)}({\bf x}-\tilde{{\bf x}})\delta(r-
\tilde{r})\;,
\label{green}
\eea
to arrive at
\be
\langle 6 \rangle^{(i)}\;=\;
\frac{2}{3}\lim_{r,\tilde{r} \to \infty}\int dr \int d^4x \int d\tilde{r}\int d^4 \tilde{x} \sqrt{-g(r,{\bf x})}\;h(r,{\bf x})\;h(r,{\bf x})\;h(r,{\bf x})
\nonumber \ee \be
\times\;
 \delta^{(4)}({\bf x}-\tilde{{\bf x}})\delta(r-
\tilde{r})
\;\widetilde{\nabla_{f}}h(\tilde{r},\tilde{{\bf x}})
\;\widetilde{\nabla^{f}}h(\tilde{r},\tilde{{\bf x}}) \;h(\tilde{r},\tilde{{\bf x}})\;
\ee
or, after integrating over the $(\tilde{r},\tilde{{\bf x}})$ coordinates,
\be
\langle 6 \rangle^{(i)}\;=\;
\frac{2}{3}\lim_{r\to \infty}\int dr \int d^4x
\sqrt{-g(r,{\bf x})}\;h(r,{\bf x})\;h(r,{\bf x})\;h(r,{\bf x})\;h(r,{\bf x})
\nonumber \ee \be \times \;
\nabla_{f} h(r,{\bf x})
\;\nabla^{f}h(r,{\bf x}) \;.
\ee
This final result is the desired six-point function,
containing the correct number of both
derivatives and gravitons. Importantly,
the process picked up a numerical factor
of $+2/3$.

It is straightforward to apply the same basic procedure to the other
two cases. Let us now suppose that both internal gravitons
are undifferentiated as in the diagram marked $(ii)$ in Fig.~4. We begin here with
\be
\langle 6 \rangle^{(ii)}\;=\;
\lim_{r,\tilde{r} \to \infty}\int dr \int d^4x \int d\tilde{r}\int d^4 \tilde{x}
\sqrt{-g(r,{\bf x})}\sqrt{-g(\tilde{r},\tilde{{\bf x}})}\;\nabla^e h(r,{\bf x})
\nonumber \ee \be
\times\;   \nabla_e
h(r,{\bf x}) \; h(r,{\bf x})
 \;\Big\langle
h(r,{\bf x}) \;h(\tilde{r},\tilde{{\bf x}}) \Big\rangle\;
\widetilde{\nabla_{f}}h(\tilde{r},\tilde{{\bf x}})
\;\widetilde{\nabla^{f}} h(\tilde{r},\tilde{{\bf x}})\;h(\tilde{r},\tilde{{\bf x}})\;,
\ee
then apply the inverse of the product rule
\be
\langle 6 \rangle^{(ii)}\;=\;
\frac{1}{2}\lim_{r,\tilde{r} \to \infty}\int dr \int d^4x  \int d\tilde{r}\int d^4 \tilde{x}
\sqrt{-g(r,{\bf x})}\sqrt{-g(\tilde{r},\tilde{{\bf x}})}
\;\nabla_e
h(r,{\bf x})
\nonumber \ee \be
\times\; \nabla^e\Big[ h(r,{\bf x})
\;h(r,{\bf x})\Big]
 \;\Big\langle
h(r,{\bf x}) \;h(\tilde{r},\tilde{{\bf x}}) \Big\rangle\;
\widetilde{\nabla_{f}}h(\tilde{r},\tilde{{\bf x}})
\;\widetilde{\nabla^{f}} h(\tilde{r},\tilde{{\bf x}})\;h(\tilde{r},\tilde{{\bf x}})
\;,
\ee
integrate by parts
\be
\langle 6 \rangle^{(ii)}\;=\;
-\frac{1}{2}\lim_{r,\tilde{r} \to \infty}\int dr \int d^4x \int d\tilde{r}\int d^4 \tilde{x}
\sqrt{-g(r,{\bf x})}\sqrt{-g(\tilde{r},\tilde{{\bf x}})} \;\nabla_e
h(r,{\bf x})
\nonumber \ee \be
\times\; h(r,{\bf x})
\;h(r,{\bf x})  \;
\nabla^e \Big\langle
h(r,{\bf x}) \;h(\tilde{r},\tilde{{\bf x}}) \Big\rangle\;
\widetilde{\nabla_{f}}h(\tilde{r},\tilde{{\bf x}})
\;\widetilde{\nabla^{f}} h(\tilde{r},\tilde{{\bf x}})\;h(\tilde{r},\tilde{{\bf x}})
\;,
\ee
reapply the inverse of the product rule
\be
\langle 6 \rangle^{(ii)}\;=\;
-\frac{1}{6}\lim_{r,\tilde{r} \to \infty}\int dr \int d^4x \int d\tilde{r}\int d^4 \tilde{x}
\sqrt{-g(r,{\bf x})}\sqrt{-g(\tilde{r},\tilde{{\bf x}})}
\nonumber \ee \be \times \;
\nabla_e\Big[h(r,{\bf x})\;h(r,{\bf x})\;h(r,{\bf x})\Big]
\;\nabla^e \Big\langle
h(r,{\bf x}) \;h(\tilde{r},\tilde{{\bf x}}) \Big\rangle\;
\widetilde{\nabla_{f}}h(\tilde{r},\tilde{{\bf x}})
\;\widetilde{\nabla^{f}} h(\tilde{r},\tilde{{\bf x}})\;h(\tilde{r},\tilde{{\bf x}})
\;,
\ee
again integrate by parts
\be
\langle 6 \rangle^{(ii)}\;=\;
\frac{1}{6}\lim_{r,\tilde{r} \to \infty}\int dr \int d^4x \int d\tilde{r}\int d^4 \tilde{x}
\sqrt{-g(r,{\bf x})}\sqrt{-g(\tilde{r},\tilde{{\bf x}})}
\;h(r,{\bf x})
\nonumber \ee \be
\times\;h(r,{\bf x})\; h(r,{\bf x}) \;
\nabla_e\nabla^e \Big\langle
h(r,{\bf x}) \;h(\tilde{r},\tilde{{\bf x}}) \Big\rangle\;
\widetilde{\nabla_{f}}h(\tilde{r},\tilde{{\bf x}})
\;\widetilde{\nabla^{f}} h(\tilde{r},\tilde{{\bf x}})\;h(\tilde{r},\tilde{{\bf x}})
\;,
\ee
call upon  Eq.~(\ref{green})
\be
\langle 6 \rangle^{(ii)}\;=\;
\frac{1}{6}\lim_{r,\tilde{r} \to \infty}\int dr \int d^4x \int d\tilde{r}\int d^4 \tilde{x}
\sqrt{-g(r,{\bf x})}\;h(r,{\bf x})\;h(r,{\bf x})\;h(r,{\bf x})
\nonumber \ee \be
\times\;
 \delta^{(4)}({\bf x}-\tilde{{\bf x}})\delta(r-
\tilde{r})
\;\widetilde{\nabla_{f}}h(\tilde{r},\tilde{{\bf x}})
\;\widetilde{\nabla^{f}}h(\tilde{r},\tilde{{\bf x}}) \;h(\tilde{r},\tilde{{\bf x}})
\;,
\ee
and  integrate over the $(\tilde{r},\tilde{{\bf x}})$ coordinates to obtain
\be
\langle 6 \rangle^{(ii)}\;=\;
\frac{1}{6}\lim_{r\to \infty}\int dr \int d^4x
\sqrt{-g(r,{\bf x})}\;h(r,{\bf x})\;h(r,{\bf x})\;h(r,{\bf x})\;h(r,{\bf x})
\nonumber \ee \be \times\;
\nabla_{f} h(r,{\bf x})
\;\nabla^{f}h(r,{\bf x}) \;.
\ee
This case comes with a numerical factor of $+1/6$.

This leaves the mixed case (one internal with a
derivative and one without) that is depicted in diagram $(iii)$ in Fig.~4. So we now consider
\be
\langle 6 \rangle^{(iii)}\;=\;
\lim_{r,\tilde{r} \to \infty}\int dr \int d^4x \int d\tilde{r}\int d^4 \tilde{x}
\sqrt{-g(r,{\bf x})}\sqrt{-g(\tilde{r},\tilde{{\bf x}})}\;h(r,{\bf x})\;h(r,{\bf x})
\nonumber \ee \be
\times\;\nabla_e
h(r,{\bf x}) \;
\Big\langle
\nabla^e h(r,{\bf x}) \;h(\tilde{r},\tilde{{\bf x}}) \Big\rangle\;
\widetilde{\nabla_{f}}h(\tilde{r},\tilde{{\bf x}})
\;\widetilde{\nabla^{f}} h(\tilde{r},\tilde{{\bf x}})\;h(\tilde{r},\tilde{{\bf x}})
\;,
\ee
then apply the inverse of the  product rule
\be
\langle 6 \rangle^{(iii)}\;=\;
\frac{1}{3}\lim_{r,\tilde{r} \to \infty}\int dr \int d^4x \int d\tilde{r}\int d^4 \tilde{x}
\sqrt{-g(r,{\bf x})}\sqrt{-g(\tilde{r},\tilde{{\bf x}})}
\nonumber \ee \be
\times\; \nabla_e\Big[h(r,{\bf x})\;h(r,{\bf x})
\;h(r,{\bf x})\Big] \;
 \nabla^e \Big\langle
h(r,{\bf x}) \;h(\tilde{r},\tilde{{\bf x}}) \Big\rangle\;
\widetilde{\nabla_{f}}h(\tilde{r},\tilde{{\bf x}})
\;\widetilde{\nabla^{f}} h(\tilde{r},\tilde{{\bf x}})\;h(\tilde{r},\tilde{{\bf x}})
\;,
\ee
integrate by parts
\be
\langle 6 \rangle^{(iii)}\;=\;
-\frac{1}{3}\lim_{r,\tilde{r} \to \infty}\int dr \int d^4x
\int d\tilde{r}\int d^4\tilde{x}
\sqrt{-g(r,{\bf x})}\sqrt{-g(\tilde{r},\tilde{{\bf x}})}\;h(r,{\bf x})
\nonumber \ee \be
\times\;h(r,{\bf x}) \;h(r,{\bf x}) \;
 \nabla_e\nabla^e \Big\langle
h(r,{\bf x}) \;h(\tilde{r},\tilde{{\bf x}}) \Big\rangle\;
\widetilde{\nabla_{f}}h(\tilde{r},\tilde{{\bf x}})
\;\widetilde{\nabla^{f}} h(\tilde{r},\tilde{{\bf x}})\;h(\tilde{r},\tilde{{\bf x}})
\;,
\ee
employ Eq.~(\ref{green})
\be
\langle 6 \rangle^{(iii)}\;=\;
-\frac{1}{3}\lim_{r,\tilde{r} \to \infty}\int dr \int d^4x \int d\tilde{r}\int d^4 \tilde{x}
\sqrt{-g(r,{\bf x})}\;h(r,{\bf x})\;h(r,{\bf x})\;h(r,{\bf x})
\nonumber \ee \be
\times\;
 \delta^{(4)}({\bf x}-\tilde{{\bf x}})\delta(r-
\tilde{r})
\;\widetilde{\nabla_{f}}h(\tilde{r},\tilde{{\bf x}})
\;\widetilde{\nabla^{f}}h(\tilde{r},\tilde{{\bf x}}) \;h(\tilde{r},\tilde{{\bf x}})
\;,
\ee
and then end by integrating
\be
\langle 6 \rangle^{(iii)}\;=\;
-\frac{1}{3}\lim_{r\to \infty}\int dr \int d^4x
\sqrt{-g(r,{\bf x})}\;h(r,{\bf x})\;h(r,{\bf x})\;h(r,{\bf x})\;h(r,{\bf x})
\nonumber \ee \be \times \;
\nabla_{f} h(r,{\bf x})
\;\nabla^{f}h(r,{\bf x}) \;.
\ee
In this case, the numerical factor is $-1/3$.

\subsubsection{Step~2 \label{step2}}

The next step is to determine the numerical factor that comes from the tensor structure. Let us first observe
that a four-point function with two derivatives has the
following form (up to additional irrelevant numerical factors and
the metric determinant):
\be
\langle 4 \rangle_{1PI}\;=\;
{\cal X}^{abcd}\nabla_e h_{ad}\nabla^e h_{bc}{\cal Y}^{pqrs}h_{ps}h_{qr}\;,
\ee
where
\be
{\cal Y}^{abcd}\;{\equiv}\;\frac{1}{2}\left[g^{ac}g^{bd}+g^{ab}g^{cd}\right]\;,
\ee
\be
{\cal X}^{abcd}\;{\equiv}\;\frac{1}{2}\left[g^{ac}g^{bd}-g^{ad}g^{bc}\right]\;.
\ee
This  structure follows from the fact that the gravitons are contracted in pairs.  Undifferentiated
gravitons originate from the expansion of the metric determinant or a contravariant metric as  symmetric contractions, whereas differentiated gravitons originate from the expansion of the Riemann tensor as anti-symmetric contractions. The transverse/traceless gauge is also relevant to the stated
form.

Other than determining the placement of the tensors ${\cal X}$ and ${\cal Y}$,
the derivatives play no role in this discussion, so we can
subsequently ignore their presence (or, equivalently,  work in momentum space).

Like before, there is an important distinction between
the choice of contracted gravitons, and we again start with
the case that both of these are differentiated.
Then the convolution of two four-points can be expressed
as
\be
\langle 6 \rangle^{(i)}\;=\;
{\cal Y}^{ijkl} h_{il}h_{jk}{\cal X}^{abcd}h_{ad}
\;\langle h_{bc}h_{ps}\rangle\; h_{qr}{\cal X}^{pqrs}h_{tw}h_{uv}
{\cal Y}^{tuvw}\;.
\ee

We next  make use of the well-known  tensor structure  of the  graviton propagator in momentum space \cite{prop1,prop2},
\be
\langle h_{ab}h_{cd}\rangle\;=\;
{\cal G}_{abcd}\;\equiv\;\frac{1}{2}\left[g_{ac}g_{bd}+g_{ad}g_{bc}
-g_{ab}g_{cd}\right]\;.
\ee
We have presented the flat-space form only for simplicity. The sole effect of the AdS curvature is to alter the coefficient of the ``trace'' (or third) term in ${\cal G}$. As can be verified, this term never contributes in any of the cases, so we will subsequently drop it. It is worth noting that boundary counter-terms (as necessary for holographic renormalization; see Section~3.1)
are non-dynamical and would
therefore similarly only alter the  trace.

Hence, the above convolution becomes
\be
\langle 6 \rangle^{(i)}\;=\;
{\cal Y}^{ijkl} h_{il}h_{jk}\Big[{\cal X}^{abcd}h_{ad}
{\cal G}_{bcps} h_{qr}{\cal X}^{pqrs}\Big]h_{tw}h_{uv}
{\cal Y}^{tuvw}\;.
\ee

Some straightforward tensor algebra then yields
\bea
\langle 6 \rangle^{(i)} & = &
\frac{1}{4}{\cal Y}^{ijkl} h_{il}h_{jk}
\Big[h_{ad}\Big(\delta^a_{\;s}\delta^d_{\;p}+\delta^a_{\;p}\delta^d_{\;s}
-2 g^{ad}g_{ps}\Big)h_{qr}{\cal X}^{prqs}\Big]
\;h_{tw}h_{uv}
{\cal Y}^{tuvw}
\nonumber \\
& = &
\frac{1}{8}{\cal Y}^{ijkl} h_{il}h_{jk}
\Big[h_{ad}\Big(g^{aq}g^{dr}+g^{ar}g^{dq}+2(d-2)g^{ad}g^{qr}\Big)h_{qr}
\Big]h_{tw}h_{uv}
{\cal Y}^{tuvw}
\nonumber \\
& =&
\frac{1}{4}{\cal Y}^{ijkl} h_{il}h_{jk}
\Big[h_{ab}h^{ab}\Big]h_{tw}h_{uv}
{\cal Y}^{tuvw}\;,
\eea
where $h^a_{\;a}=0$ has been used in the last line
and $d$ is meant as the dimensionality of the graviton sector.

Notice that only the initial partners of the internal
modes are involved in this process, with the rest of the
gravitons simply ``along for the ride''.
Also, the factor of $1/4$ should be kept in mind.

The other two cases follow along similar lines.
If the two internal gravitons are both undifferentiated,
then
\bea
\langle 6 \rangle^{(ii)} & = &
{\cal X}^{ijkl} h_{il}h_{jk}\Big[{\cal Y}^{abcd}h_{ad}
{\cal G}_{bcps} h_{qr}{\cal Y}^{pqrs}\Big]h_{tw}h_{uv}
{\cal X}^{tuvw}
\nonumber \\ & = &
\frac{1}{2}{\cal X}^{ijkl} h_{il}h_{jk}
\Big[h_{ad}\Big(\delta^a_{\;s}\delta^d_{\;p}+\delta^a_{\;p}\delta^d_{\;s}
\Big)h_{qr}\;{\cal Y}^{prqs}\Big]
h_{tw}h_{uv}
{\cal X}^{tuvw}
\nonumber \\ & = &
\frac{1}{2}{\cal X}^{ijkl} h_{il}h_{jk}
\Big[h_{ad}\Big(g^{aq}g^{dr}+g^{ar}g^{dq}
\Big)
h_{qr}\Big]h_{tw}h_{uv}
{\cal X}^{tuvw}
\nonumber \\ & = &
1\cdot{\cal X}^{ijkl} h_{il}h_{jk}
\Big[h_{ab}h^{ab}\Big]h_{tw}h_{uv}
{\cal X}^{tuvw}\;.
\eea

And, when the internal gravitons are ``mixed'',
\bea
\langle 6 \rangle^{(iii)} & = &
{\cal Y}^{ijkl} h_{il}h_{jk}\Big[{\cal X}^{abcd}h_{ad}
{\cal G}_{bcps}  h_{qr}{\cal Y}^{pqrs}\Big]h_{tw}h_{uv}
{\cal X}^{tuvw}
\nonumber \\ & = &
\frac{1}{4}{\cal Y}^{ijkl} h_{il}h_{jk}
\Big[h_{ad}\Big(\delta^a_{\;s}\delta^d_{\;p}+\delta^a_{\;p}\delta^d_{\;s}
-2 g^{ad}g_{ps}\Big)h_{qr}\;{\cal Y}^{prqs}
\Big]h_{tw}h_{uv}
{\cal X}^{tuvw}
\nonumber \\ & = &
\frac{1}{4}{\cal Y}^{ijkl} h_{il}h_{jk}
\Big[h_{ad}\Big(g^{aq}g^{dr}+g^{ar}g^{dq} -2 g^{ad}g^{qr}\Big)
h_{qr}\Big]h_{tw}h_{uv}
{\cal X}^{tuvw}
\nonumber \\ & = &
\frac{1}{2}{\cal Y}^{ijkl} h_{il}h_{jk}
\Big[h_{ab}h^{ab}\Big]h_{tw}h_{uv}
{\cal X}^{tuvw}\;.
\eea

Again take note of the  numerical factors; respectively,
1 and $1/2$.

\subsubsection{Putting it all together}

Let us first recall the relative weight  factor of $\;c_6=10\;$ that was
obtained from Feynman combinatorics.
One can see that half of this or  a factor of 5
should be attributed to the mixed case, whereas a factor of
5/2 should be assigned  to  either case  with a  matched pair.

We are finally in a position to determine the  net relative weight of the 1PR contribution to the Einstein theory six-point function. The contribution of any one case is given by the product of its numerical factors for Steps~1 and~2 times the above combinatoric factor. The three outcomes are then summed together.

When the two internal gravitons are both differentiated,\\
$\;\frac{2}{3}\times\frac{1}{4}\times\frac{5}{2}=\frac{5}{12}\;$, \\
both undifferentiated, \\
$\;\frac{1}{6}\times 1\times\frac{5}{2}=\frac{5}{12}\;$, \\
and when there is one of each, \\
$\;-\frac{1}{3}\times\frac{1}{2}\times{5}=-\frac{5}{6}\;$, \\
for a total of $\frac{5}{12}+
\frac{5}{12}-\frac{5}{6}$ or zero~!

\subsection{...for Gauss--Bonnet gravity}

\subsubsection{Initial considerations}

Let us now consider
the  leading-order Gauss--Bonnet correction to the preceding 1PR calculation. By insisting on the high-momentum regime  and  working to order $\epsilon$, we must
 have  either $\langle hhhh\rangle_{\epsilon^0} \langle hh\rangle^{-1}_{\epsilon^0}
\langle hhhh\rangle_{\epsilon^1}$ or $\langle hhhh\rangle_{\epsilon^1} \langle hh\rangle^{-1}_{\epsilon^0}
\langle hhhh\rangle_{\epsilon^0}$. Hence, there will be six derivatives in
total  and,  since the process of
contraction reduces this by two,  the end result is Gauss--Bonnet's
$s^2$ signature. Otherwise,
the general procedure closely follows the previous Einstein theory calculations.

\subsubsection{Step 1}

\begin{figure}[t]
\vspace{-0.5in}
\scalebox{.45}{\includegraphics[angle=-90]{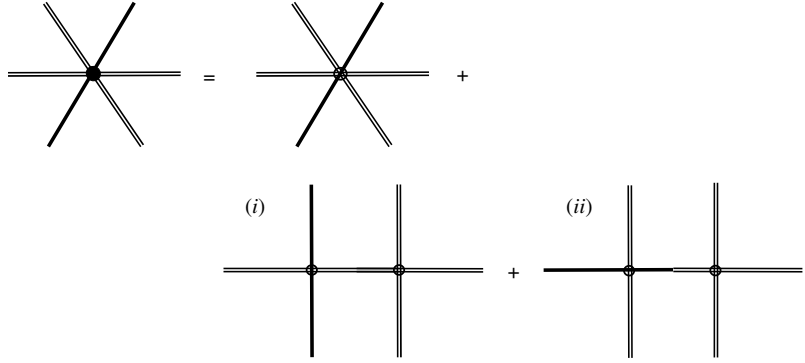}}
\caption{Six-point function at order {$\epsilon$} resulting from the GB term.}
\end{figure}

At order $\epsilon^1$, there are only two possible cases,
as at least one internal graviton must be differentiated. These are depicted in diagrams $(i)$ and $(ii)$ in Fig.~5.
Suppose that both internal gravitons are differentiated, then
the starting point is
\be
\langle 6 \rangle^{(i),\epsilon}\;=\;
\lim_{r,\tilde{r} \to \infty}\int dr \int d^4x \int d\tilde{r}\int d^4 \tilde{x}
\sqrt{-g(r,{\bf x})}\sqrt{-g(\tilde{r},\tilde{{\bf x}})}\;\nabla_d h(r,{\bf x})
\nonumber \ee \be
\times\; \nabla^d h(r,{\bf x}) \;\nabla_e h(r,{\bf x}) \;
\Big\langle
\nabla^e h(r,{\bf x})\;\widetilde{\nabla^{f}} h(\tilde{r},\tilde{{\bf x}}) \Big\rangle\;
\widetilde{\nabla_{f}}h(\tilde{r},\tilde{{\bf x}})
 \;h(\tilde{r},\tilde{{\bf x}})\;h(\tilde{r},\tilde{{\bf x}}) \;.
\ee

Proceeding  just like before, we
first apply the inverse of the product rule
\be
\langle 6 \rangle^{(i),\epsilon}\;=\;
\frac{1}{3}\lim_{r,\tilde{r} \to \infty}\int dr \int d^4x\int d\tilde{r}\int d^4 \tilde{x}
\sqrt{-g(r,{\bf x})}\sqrt{-g(\tilde{r},\tilde{{\bf x}})}
\nonumber \ee \be
\times\; \nabla_e\Big[\nabla_d h(r,{\bf x})
\;\nabla^d h(r,{\bf x}) \;h(r,{\bf x})\Big] \;
\nabla^e \Big\langle
h(r,{\bf x})\;\widetilde{\nabla^{f}} h(\tilde{r},\tilde{{\bf x}}) \Big\rangle\;
\widetilde{\nabla_{f}}h(\tilde{r},\tilde{{\bf x}})
 \;h(\tilde{r},\tilde{{\bf x}})\;h(\tilde{r},\tilde{{\bf x}})
\;,
\ee
integrate by parts
\be
\langle 6 \rangle^{(i),\epsilon}\;=\;
-\frac{1}{3}\lim_{r,\tilde{r} \to \infty}\int dr \int d^4x \int d\tilde{r}\int d^4 \tilde{x}
\sqrt{-g(r,{\bf x})}\sqrt{-g(\tilde{r},\tilde{{\bf x}})} \;h(r,{\bf x})
\nonumber \ee \be
\times\; \nabla_d h(r,{\bf x})
\;\nabla^d h(r,{\bf x})  \; \nabla_e\nabla^e\Big\langle
 h(r,{\bf x})\;\widetilde{\nabla^{f}} h(\tilde{r},\tilde{{\bf x}}) \Big\rangle\;
\widetilde{\nabla_{f}}h(\tilde{r},\tilde{{\bf x}})
 \;h(\tilde{r},\tilde{{\bf x}})\;h(\tilde{r},\tilde{{\bf x}})
\;,
\ee
reapply the inverse product rule
\be
\langle 6 \rangle^{(i),\epsilon}\;=\;
\frac{1}{3}\lim_{r,\tilde{r} \to \infty}\int dr \int d^4x \int d\tilde{r}\int d^4 \tilde{x}
\sqrt{-g(r,{\bf x})}\sqrt{-g(\tilde{r},\tilde{{\bf x}})} \;h(r,{\bf x})
\nonumber \ee \be
\times\; \nabla_d h(r,{\bf x})
\;\nabla^d h(r,{\bf x})\; \nabla_e\nabla^e\Big\langle
 h(r,{\bf x}) \;h(\tilde{r},\tilde{{\bf x}}) \Big\rangle\;
\widetilde{\nabla^{f}}\Big[\widetilde{\nabla_{f}}h(\tilde{r},\tilde{{\bf x}})
 \;h(\tilde{r},\tilde{{\bf x}})\;h(\tilde{r},\tilde{{\bf x}})\Big]\;,
\ee
again integrate by parts
\be
\langle 6 \rangle^{(i),\epsilon}\;=\;
\frac{2}{3}\lim_{r,\tilde{r} \to \infty}\int dr \int d^4x \int d\tilde{r}\int d^4\tilde{x}
\sqrt{-g(r,{\bf x})}\sqrt{-g(\tilde{r},\tilde{{\bf x}})} \;h(r,{\bf x})
\nonumber \ee \be
\times\; \nabla_d h(r,{\bf x})
\;\nabla^d h(r,{\bf x}) \;
\nabla_e\nabla^e\Big\langle
 h(r,{\bf x}) \;h(\tilde{r},\tilde{{\bf x}}) \Big\rangle\;
\widetilde{\nabla^{f}}  h(\tilde{r},\tilde{{\bf x}})    \;\widetilde{\nabla_{f}}
 h(\tilde{r},\tilde{{\bf x}})\;h(\tilde{r},\tilde{{\bf x}})
\;,
\ee
incorporate the Green's function identity~(\ref{green})
\be
\langle 6 \rangle^{(i),\epsilon}\;=\;
\frac{2}{3}\lim_{r,\tilde{r} \to \infty}\int dr \int d^4x \int d\tilde{r}\int d^4 \tilde{x}
\sqrt{-g(r,{\bf x})}\;h(r,{\bf x}) \;\nabla_d h(r,{\bf x}) \nonumber \ee \be
\times\;  \nabla^d h(r,{\bf x}) \;
 \delta^{(4)}({\bf x}-\tilde{{\bf x}})\delta(r-
\tilde{r})
\;\widetilde{\nabla_{f}}h(\tilde{r},\tilde{{\bf x}})
\;\widetilde{\nabla^{f}}h(\tilde{r},\tilde{{\bf x}}) \;h(\tilde{r},\tilde{{\bf x}})
\;,
\ee
and finish by integrating over the $(\tilde{r},\tilde{{\bf x}})$ coordinates
\be
\langle 6 \rangle^{(i),\epsilon}\;=\;
\frac{2}{3}\lim_{r\to \infty}\int dr \int d^4x
\sqrt{-g(r,{\bf x})}\;h(r,{\bf x}) \;\nabla_d h(r,{\bf x})\;\nabla_d h(r,{\bf x})
 \;h(r,{\bf x}) \nonumber \ee \be \times \;
\nabla_{f} h(r,{\bf x})
\;\nabla^{f}h(r,{\bf x}) \;,
\ee
which is the expected $s^2$ form.
As usual, take note of the factor $+2/3$.

Suppose now that only one  internal graviton is differentiated,
then
\be
\langle 6 \rangle^{(ii),\epsilon}\;=\;
\lim_{r,\tilde{r} \to \infty}\int dr \int d^4x \int d\tilde{r}\int d^4\tilde{x}
\sqrt{-g(r,{\bf x})}\sqrt{-g(\tilde{r},\tilde{{\bf x}})}\;\nabla_d h(r,{\bf x})
\nonumber \ee \be
\times\; \nabla^d h(r,{\bf x}) \;\nabla_e h(r,{\bf x}) \;
 \Big\langle
\nabla^e h(r,{\bf x}) \;h(\tilde{r},\tilde{{\bf x}}) \Big\rangle\;  \;h(\tilde{r},\tilde{{\bf x}})
\;\widetilde{\nabla^{f}}h(\tilde{r},\tilde{{\bf x}})
\;\widetilde{\nabla_{f}} h(\tilde{r},\tilde{{\bf x}})
\;.
\ee
Applying the inverse product rule
\be
\langle 6 \rangle^{(ii),\epsilon}\;=\;
\frac{1}{3}\lim_{r,\tilde{r} \to \infty}\int dr \int d^4x \int d\tilde{r}\int d^4 \tilde{x}
\sqrt{-g(r,{\bf x})}\sqrt{-g(\tilde{r},\tilde{{\bf x}})}
\nonumber \ee \be
\times\; \nabla_e\Big[\nabla_d h(r,{\bf x})
\;\nabla^d h(r,{\bf x}) \;h(r,{\bf x}) \Big] \;
 \nabla^e \Big\langle
 h(r,{\bf x}) \;h(\tilde{r},\tilde{{\bf x}}) \Big\rangle\;  h(\tilde{r},\tilde{{\bf x}})
\;\widetilde{\nabla^{f}}h(\tilde{r},\tilde{{\bf x}})
\;\widetilde{\nabla_{f}} h(\tilde{r},\tilde{{\bf x}})
\;,
\ee
integrating by parts
\be
\langle 6 \rangle^{(ii),\epsilon}\;=\;
-\frac{1}{3}\lim_{r,\tilde{r} \to \infty}\int dr \int d^4x \int d\tilde{r}\int d^4 \tilde{x}
\sqrt{-g(r,{\bf x})}\sqrt{-g(\tilde{r},\tilde{{\bf x}})} \;\nabla_d h(r,{\bf x})
\nonumber \ee \be
\times\; \nabla^d h(r,{\bf x}) \;
 h(r,{\bf x}) \; \nabla_e\nabla^e \Big\langle
 h(r,{\bf x}) \;h(\tilde{r},\tilde{{\bf x}}) \Big\rangle\;  h(\tilde{r},\tilde{{\bf x}})
\;\widetilde{\nabla^{f}}h(\tilde{r},\tilde{{\bf x}})
\;\widetilde{\nabla_{f}} h(\tilde{r},\tilde{{\bf x}})
\;,
\ee
utilizing Eq.~(\ref{green})
\be
\langle 6 \rangle^{(ii),\epsilon}\;=\;
-\frac{1}{3}\lim_{r,\tilde{r} \to \infty}\int dr
\int d^4x \int d\tilde{r}\int d^4 \tilde{x}
\sqrt{-g(r,{\bf x})}\;h(r,{\bf x}) \;\nabla_d h(r,{\bf x})
\nonumber \ee \be
\times\;  \nabla^d h(r,{\bf x}) \;
 \delta^{(4)}({\bf x}-\tilde{{\bf x}})\delta(r-
\tilde{r})
\;\widetilde{\nabla_{f}}h(\tilde{r},\tilde{{\bf x}})
\;\widetilde{\nabla^{f}}h(\tilde{r},\tilde{{\bf x}}) \;h(\tilde{r},\tilde{{\bf x}})
\;,
\ee
and then integrating over ${\tilde r}$ and ${\tilde{{\bf x}}}$, we have
\be
\langle 6 \rangle^{(ii),\epsilon}\;=\;
-\frac{1}{3}\lim_{r\to \infty}\int dr \int d^4x
\sqrt{-g(r,{\bf x})}\;h(r,{\bf x}) \;\nabla_d h(r,{\bf x})\;\nabla_d h(r,{\bf x})
 \;h(r,{\bf x})
\nonumber \ee \be \times \;
\nabla_{f} h(r,{\bf x})
\;\nabla^{f}h(r,{\bf x})
\;.
\ee
This time the relevant factor is  $-1/3$.

\subsubsection{Step 2 and beyond}

The results of Step~2 at order $\epsilon$ are identical to those of the Einstein theory calculation.
The salient points are that
the order-$\epsilon$ limit constrains all the background
structure (${\cal X}$, ${\cal Y}$, ${\cal G}$) to be fixed at order
$\epsilon^0$~\footnote{One might worry about starting with Einstein gravity and
then inserting  $\epsilon$-corrected forms for
${\cal X}$, ${\cal Y}$, ${\cal G}$, but this would violate
the high-momentum condition that $\epsilon$-order amplitudes contain
four derivatives.} and  the additional differentiated gravitons are only ``along for the ride''.
And so  there is a resulting factor of $1/4$
when the two internal gravitons are differentiated and a factor
of $1/2$ when only one is differentiated.

As for Feynman combinatorics,
we again recall the relative weight factor $\;c_6=10\;$, but there
is now an extra factor of 2 because
of the two ways of placing the $\epsilon$.
It is easy to see that this number should  be distributed
evenly between the two viable cases, leading to a common factor of
$\;10\cdot2/2=10\;$.

Finally, let us put it all together:\\
When the two internal gravitons are both differentiated,\\
$\;\frac{2}{3}\times\frac{1}{4}\times{10}=\frac{5}{3}\;$, \\
and when there is one of each, \\
$\;-\frac{1}{3}\times\frac{1}{2}\times{10}=-\frac{5}{3}\;$; \\
so that, even at order $\epsilon$, the net contribution
conspires to vanish.

\section{General 1PR functions...}

We will next show that the previously observed cancelation persists for any diagram with a single internal line.

\subsection{...for Einstein gravity \label{pq}}

We begin with pure Einstein gravity and
consider a  1PR function
that is formed by  convolving a  single pair of amplitudes but is otherwise general,
\be
\langle h^{2n}\rangle_{{2p-2q}} \;=\; \frac{\langle h^{2p}\rangle_{1PI}
\langle h^{2q}\rangle_{1PI}}{\langle hh\rangle}\;,
\ee
where $\;2p,2q\geq 4\;$ and $\;2n=2p+2q-2\;$. See Fig.~6.

\begin{figure}[t]
\vspace{-2.5in}
\scalebox{.65}{\includegraphics[angle=-90]{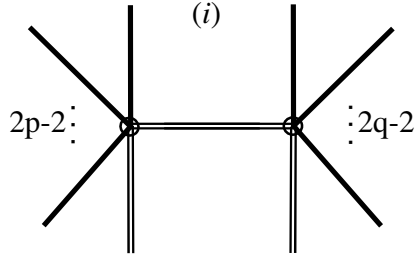}}
\caption{A reducible $2p-2q$ diagram with $2p+2q=2n$. This is one of
the three
1PR diagrams contributing to the $2n$-point function.}
\end{figure}

It is a straigthforward exercise to generalize the steps of
the previous $\;2n=6\;$ analysis. This is because
the number of external gravitons only has bearing  on the use
of the product rule (and its inverse) in Step~1 and the
distribution of the Feynman combinatoric factor.
Nothing has changed regarding Step~2.

Let us show exactly how this works, starting with Step~1 for
 the case of
of both internal gravitons being differentiated,
\be
\langle 2n \rangle^{(i)}_{ {2p-2q}}\;=\;
 \lim_{r,\tilde{r} \to \infty}\int dr \int d^4x \int d\tilde{r}\int d^4 \tilde{x}
\sqrt{-g(r,{\bf x})}\sqrt{-g(\tilde{r},\tilde{{\bf x}})}\big[h(r,{\bf x})\big]^{2p-2}
\nonumber \ee \be
\times\; \nabla_e h(r,{\bf x}) \;
\Big\langle
\nabla^e h(r,{\bf x})\;\widetilde{\nabla^{f}} h(\tilde{r},\tilde{{\bf x}})
\Big\rangle\;
\widetilde{\nabla_{f}}h(\tilde{r},\tilde{{\bf x}})
 \big[h(\tilde{r},\tilde{{\bf x}})\big]^{2q-2} \;.
\ee

The same  manipulations as in $\S$~\ref{stepone} can be applied. We first integrate by parts
\be
\langle 2n \rangle^{(i)}_{ {2p-2q}}\;=\;-\lim_{r,\tilde{r} \to \infty}\int dr \int d^4x \int d\tilde{r}\int d^4 \tilde{x}
\sqrt{-g(r,{\bf x})}\sqrt{-g(\tilde{r},\tilde{{\bf x}})}\big[h(r,{\bf x})\big]^{2p-2}
\nonumber \ee
\be
\times\; \nabla_e h(r,{\bf x}) \;
 \Big\langle
\nabla^e h(r,{\bf x}) \;h(\tilde{r},\tilde{{\bf x}}) \Big\rangle\;
\widetilde{\nabla_{f}}h(\tilde{r},\tilde{{\bf x}})
\;\widetilde{\nabla^{f}}\Big(\big[h(\tilde{r},\tilde{{\bf x}})\big]^{2q-2}\Big)
\;,
\ee
then apply the product rule
\be
\langle 2n \rangle^{(i)}_{ {2p-2q}}\;=\;-(2q-2)\lim_{r,\tilde{r} \to \infty}\int dr \int d^4x \int d\tilde{r}\int d^4 \tilde{x}
\sqrt{-g(r,{\bf x})}\sqrt{-g(\tilde{r},\tilde{{\bf x}})}
\nonumber \ee
\be
\times\;  \big[h(r,{\bf x})\big]^{2p-2}\;\nabla_e h(r,{\bf x}) \;
\Big\langle
\nabla^e h(r,{\bf x}) \;h(\tilde{r},\tilde{{\bf x}}) \Big\rangle\;
\widetilde{\nabla_{f}}h(\tilde{r},\tilde{{\bf x}})
\;\widetilde{\nabla^{f}}h(\tilde{r},\tilde{{\bf x}}) \big[h(\tilde{r},\tilde{{\bf x}})\big]^{2q-3}
\;,
\ee
next apply the inverse of the  product rule
\be
\langle 2n \rangle^{(i)}_{ {2p-2q}}\;=\;-\left(\frac{2q-2}{2p-1}\right)\lim_{r,\tilde{r} \to \infty}\int dr \int d^4x \int d\tilde{r}\int d^4 \tilde{x}
\sqrt{-g(r,{\bf x})}\sqrt{-g(\tilde{r},\tilde{{\bf x}})}
\nonumber \ee \be
\times\; \nabla_e\Big(\big[h(r,{\bf x})\big]^{2p-1}\Big) \;
 \Big\langle
\nabla^e h(r,{\bf x}) \;h(\tilde{r},\tilde{{\bf x}}) \Big\rangle\;
\widetilde{\nabla_{f}}h(\tilde{r},\tilde{{\bf x}})
\;\widetilde{\nabla^{f}}h(\tilde{r},\tilde{{\bf x}}) \big[h(\tilde{r},\tilde{{\bf x}})\big]^{2q-3}
\;,
\ee
integrate by parts for a second time
\be
\langle 2n \rangle^{(i)}_{ {2p-2q}}\;=\;\left(\frac{2q-2}{2p-1}\right)\lim_{r,\tilde{r} \to \infty}\int dr \int d^4x \int d\tilde{r}\int d^4 \tilde{x}
\sqrt{-g(r,{\bf x})}\sqrt{-g(\tilde{r},\tilde{{\bf x}})}
\nonumber \ee \be
\times\;
\big[h(r,{\bf x})\big]^{2p-1} \;
\nabla_e\nabla^e \Big\langle h(r,{\bf x}) \;h(\tilde{r},\tilde{{\bf x}}) \Big\rangle\;
\widetilde{\nabla_{f}}h(\tilde{r},\tilde{{\bf x}})
\;\widetilde{\nabla^{f}}h(\tilde{r},\tilde{{\bf x}}) \big[h(\tilde{r},\tilde{{\bf x}})\big]^{2q-3}
\;,
\ee
use the Green's function identity~(\ref{green})
\be
\langle 2n \rangle^{(i)}_{ {2p-2q}}\;=\;\left(\frac{2q-2}{2p-1}\right)\lim_{r,\tilde{r} \to \infty}\int dr \int d^4x \int d\tilde{r}\int d^4 \tilde{x}
\sqrt{-g(r,{\bf x})}\big[h(r,{\bf x})\big]^{2p-1}
\nonumber \ee \be
\times\;
 \delta^{(4)}({\bf x}-\tilde{{\bf x}})\delta(r-
\tilde{r})
\;\widetilde{\nabla_{f}}h(\tilde{r},\tilde{{\bf x}})
\;\widetilde{\nabla^{f}}h(\tilde{r},\tilde{{\bf x}})\big[h(\tilde{r},\tilde{{\bf x}})\big]^{2q-3}\;
\;,
\ee
and end by integrating over the $(\tilde{r},\tilde{{\bf x}})$ coordinates
\be
\langle 2n \rangle^{(i)}_{ {2p-2q}}\;=\;\left(\frac{2q-2}{2p-1}\right)\lim_{r\to \infty}\int dr \int d^4x
\sqrt{-g(r,{\bf x})}\big[h(r,{\bf x})\big]^{2(n-1)}
\nonumber \ee \be \times \;
\nabla_{f} h(r,{\bf x})
\;\nabla^{f}h(r,{\bf x}) \;.
\ee

The resulting 1PR contribution appears to picks up a factor of $\frac{2q-2}{2p-1}$. But, on the other hand,  had we reversed the roles played by $p$ and $q$, the factor would rather be  $\frac{2p-2}{2q-1}$. This is  only an apparent ambiguity, as we have not yet accounted for the fact that the external gravitons should be  symmetrized. This symmetrization can be implemented by weighing the choice to integrate first over $r,{\bf x}$ by $\frac{2p-1}{2p+2q-2}$ and the choice to integrate first over $\tilde{r},\tilde{{\bf x}}$ by $\frac{2q-1}{2p+2q-2}\;$; with the weights determined by the relative numbers of external gravitons (this rule will be further motivated below). Hence, the correct factor becomes $\;\frac{2p-1}{2p+2q-2}\cdot\frac{2q-2}{2p-1}+\frac{2q-1}{2p+2q-2}\cdot\frac{2p-2}{2q-1}=\frac{p+q-2}{p+q-1}\;$.

Similarly, the case
of two undifferentiated gravitons picks up a factor of
$\;\frac{2p-1}{2p+2q-2}\cdot\frac{1}{(2p-1)(2p-2)}+\frac{2q-1}{2p+2q-2}\cdot\frac{1}{(2q-1)(2q-2)}=
\frac{1}{4}\frac{p+q-2}{(p+q-1)(p-1)(q-1)}\;$.
 The mixed case is a bit  different than before,
as the identity of the differentiated  internal graviton
becomes important.~\footnote{This importance is  illusionary,
as one can see by  summing the outcomes for these two sub-cases at any step.}
 If the derivative is carried by the internal graviton  ``on the left'' (or that from the $2p$-function),
then the factor goes according to
$\;\frac{2p-1}{2p+2q-2}\cdot\left[-\frac{1}{2p-1}\right]+\frac{2q-1}{2p+2q-2}\cdot\left[-\frac{2p-2}{(2q-1)(2q-2)}
\right]= -\frac{1}{2(q-1)}\frac{p+q-2}{p+q-1}\;$.
If the derivative is rather ``on the right'', then
$\;-\frac{1}{2(p-1)}\frac{p+q-2}{p+q-1}\;$.

As mentioned, Step~2 gives the same outcomes as before
(factors of 1/4, 1/2 and 1 for the three cases, respectively).
As for Feynman combinatorics, it is a simple matter to generalize
the six-point case;  as in Section~\ref{combo}.
Here, the Feynman relative weight is
$\;2p\cdot 2q\cdot (2p+2q-2)!/(2p)!(2q)!=4pqw\;$ with $\;w\equiv(2n)!/(2p)!(2q)!\;$.~\footnote{If $\;p=q\;$, there is
an extra factor of $1/2!$ coming from the expansion of the
exponent.}

Next, consider that there is either a $1/p$ or $1/q$ chance that
any given internal graviton is differentiated and, respectively,
a $(p-1)/p$ and $(q-1)/q$ chance that it is not.
Meaning that the  weight factor
should be distributed
in the following way: $\;\frac{1}{p}\cdot\frac{1}{q}\cdot
4pqw= 4w\;$
for two differentiated internals,
  $\;\frac{p-1}{p}\cdot\frac{q-1}{q}\cdot 4pqw
=4(p-1)(q-1)w\;$
for two
undifferentiated internals,
 $\;\frac{1}{p}\cdot\frac{q-1}{q}\cdot 4pqw
=4(q-1)w\;$ when there is one  differentiated internal on the left
and
 $\;\frac{p-1}{p}\cdot\frac{1}{q}\cdot 4pqw
=4(p-1)w\;$ when there is one differentiated internal  on the right.

Let us now combine the various factors:\\
When the two internal gravitons are both differentiated,\\
$\;\frac{p+q-2}{p+q-1} \times\frac{1}{4}\times 4w =
w\frac{p+q-2}{p+q-1}\;$, \\
when the two internal gravitons are both undifferentiated,\\
$\;\frac{1}{4}\frac{p+q-2}{(p+q-1)(p-1)(q-1)}
\times 1 \times 4(p-1)(q-1)w =
w\frac{p+q-2}{p+q-1}\;$, \\
when only the one on the left is differentiated, \\
$\;-\frac{1}{2(q-1)}\frac{p+q-2}{p+q-1}
\times\frac{1}{2}\times 4(q-1)w=
-w\frac{p+q-2}{p+q-1}\;$ \\
and when only the one on the right is differentiated, \\
$\;-\frac{1}{2(p-1)}\frac{p+q-2}{p+q-1}
\times\frac{1}{2}\times 4(p-1)w=
-w\frac{p+q-2}{p+q-1}\;$; \\
all of which sums to  zero, irrespective of the values
of $p$ and $q$.

Notice that any of the four types of diagrams
makes a contribution that goes, up to $\pm w$, as
$\;\frac{p+q-2}{p+q-1}=\frac{2n-4}{2n}\;$.
This is relevant because it adds credence to our rule for symmetrizing
the external gravitons. The point is that, from a momentum-space perspective,
which gravitons are on which side is completely irrelevant. Hence,
the net contributions (prior to the final summation) can
 only depend on the { total} number of external gravitons $2n$,
exactly what is found here.~\footnote{The factor $w$ can
and does carry information about $2p$ and $2q$ because it
is keeping track of how the external gravitons are drawn out of
the Lagrangian.}

 Indeed, given that the symmetrization  process must
be unambiguous, reduce to the correct procedure when $\;p=q\;$
and intermediary results can only depend on the sum $2p+2q$, it seems likely
that our rule is  the unique one.

\subsection{...for Gauss--Bonnet gravity}

We now show that the same type of cancellation persists for the
order-$\epsilon$ contributions from Gauss--Bonnet gravity.
Step~1 (and of course Step~2) works the same as for the Einstein gravity calculation, as
the formalism automatically handles the situations when two undifferentiated
internal gravitons is not viable ($p\;{\rm or}\;q=2)$.
What is now different is the distribution of the Feynman combinatoric factor,
as discussed next.

Let us  assume  for the time being  that
 $\epsilon$ goes with the $2p$-point function.
With $\epsilon$ fixed, the relative  weight factor  is  the same
as it was for pure Einstein gravity, $4pqw$ . What is different is the
fraction of internal gravitons that are differentiated;
this being $\;2/p\;$ on one side but $\;1/q\;$ on  the other.
The fraction that is undifferentiated is, respectively,
 $(p-2)/p$ and $(q-1)/q$.
Meaning that the relative weight factor
should now be redistributed according to
$\;\frac{2}{p}\cdot\frac{1}{q}\cdot
4pqw = 8w\;$
for two differentiated internals,
  $\;\frac{p-2}{p}\cdot\frac{q-1}{q}\cdot 4pqw
= 4(p-2)(q-1)w\;$
for two
undifferentiated internals,
$\;\frac{2}{p}\cdot\frac{q-1}{q}
\cdot 4pqw
=8(q-1)w\;$
for only one differentiated internal on the left and
$\;\frac{p-2}{p}\cdot\frac{1}{q}
\cdot 4pqw
=4(p-2)w\;$
for only one differentiated internal on the right.

The combined factors then go as follows: \\
When the two internal gravitons are both differentiated,\\
$\;\frac{p+q-2}{p+q-1}
\times\frac{1}{4}\times 8w =
2w\frac{p+q-2}{p+q-1}\;$, \\
when the two internal gravitons are both undifferentiated,\\
$\;\frac{1}{4}\frac{p+q-2}{(p+q-1)(p-1)(q-1)}
\times 1 \times 4(p-2)(q-1)w =
w\frac{p-2}{p-1}\frac{p+q-2}{p+q-1}\;$, \\
when only the one on the left is differentiated, \\
$\;-\frac{1}{2(q-1)}\frac{p+q-2}{p+q-1}\times\frac{1}{2}\times 8(q-1)w=
-2w\frac{p+q-2}{p+q-1}\;$ \\
and when only the one on the right is differentiated, \\
$\;-\frac{1}{2(p-1)}\frac{p+q-2}{p+q-1}\times\frac{1}{2}\times 4(p-2)w=
-w\frac{p-2}{p-1}\frac{p+q-2}{p+q-1}\;$.~\footnote{Our previous claim that intermediary results
should only depend
on the total number of external gravitons need not and does not apply at this stage of
the Gauss--Bonnet calculation.
This is because the placement of $\epsilon$ has distinguished between the two sides.}
 \\

The above sums to zero,
as must also be the case when
the $2q$-point function carries the $\epsilon$.
So that, even at order $\epsilon$, the same
cancelations  persist.

\section{The case against 1PR functions}

The results of the previous subsection are enough to tell us
that the 1PI and connected four- and six-point functions are equal up to order $\epsilon$.

What about higher-point functions? Of course, their 1PR contributions involve more complicated arrangements than so-far dealt with. For instance, the connected twelve-point function can be formed out of (a) three four-point functions plus a six-point function, (b) two six-point functions plus a four-point function and
(c) two four-point functions plus an eight-point function, as well
as (d) contributions  with a single internal line.
These combinations are shown in Fig.~7. Such calculations are exponentially more involved than those of the $2p$-$2q$ type. Yet, it can be argued on general grounds that any of these 1PR contributions must similarly vanish.

\begin{figure}[t]
\vspace{-0.55in}\hspace{-1.25in}
\scalebox{.65}{\includegraphics[angle=-90,clip=true,trim=0 0 0 0] {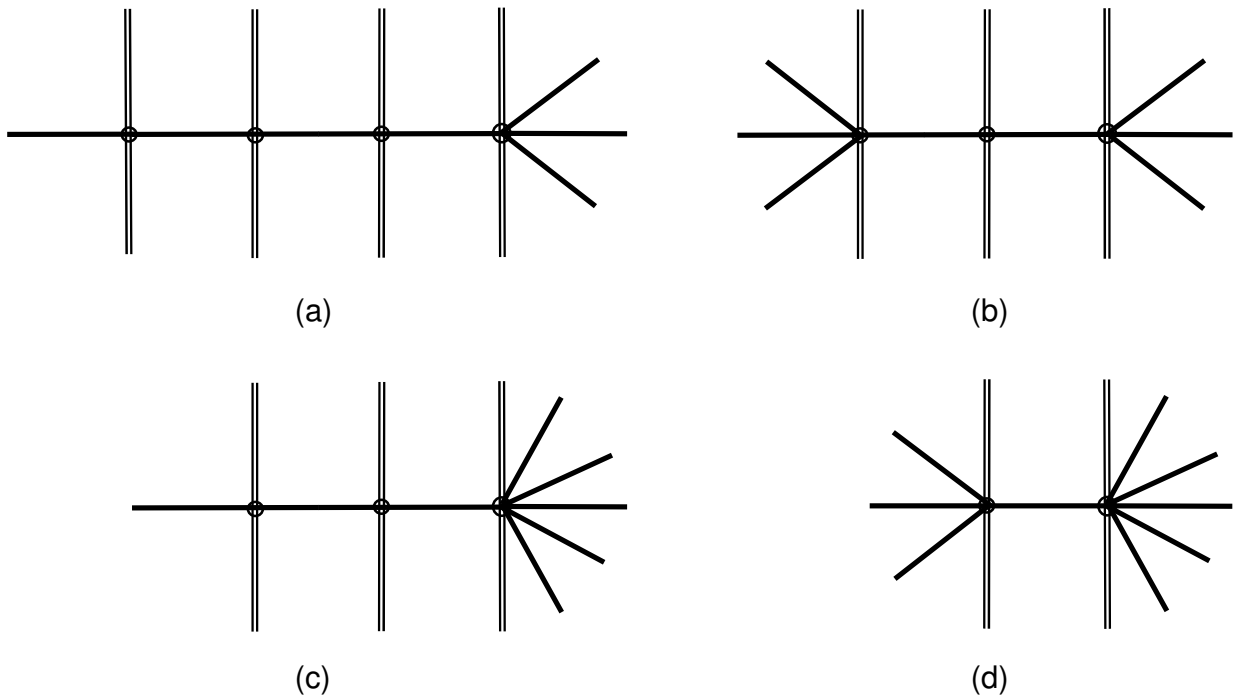}}
\caption{Four examples of reducible contributions to the 12-point function.}
\end{figure}

First, consider that the generating functionals for the connected and 1PI amplitudes are the same up to
a Legendre transformation. In either case, the generating functional contains an exponentiated action
(the physical action for the connected generating functional and the effective action for
the 1PI generating functional)
and one  can (in principle) read off the  coefficients of the amplitudes
from a Taylor expansion of the  exponential.
For a  general theory, there is no reason for the connected and 1PI coefficients to be equal. Einstein gravity in the high-momentum regime is, however, different. General covariance and  the restriction to two derivatives  conspire to severely constrain the form of either expansion  \cite{Hof,cavepeeps}. Essentially, all the relevant  terms in both expansions depend on a single number, the numerical coefficient of the Ricci scalar. As the Legendre transformation has no bearing on this number,  the two expansions  have to be equal.

We can clarify the above argument  as follows:
The generating functional for the Einstein theory connected
amplitudes can be cast in the standard form (indices are suppressed)
\be
i{\cal W}[J] \;=\ \ln\Bigg[\int {\cal D}[h]\;
 e^{i\int dx^5  \left[\sqrt{-g}\left(a{\cal R}+b\Lambda\right)+Jh\right]}\Bigg]\;,
 \ee
where ${\cal R}$ is the Ricci scalar, $\Lambda$ is the cosmological constant,
$J$ is an external source, $a,b$ are numbers and (here only) $x$ represents  all five spacetime
coordinates.
The  connected $2n$-point function is then obtained from the relation
 \be
\langle h(x_1)\cdots h(x_{2n}) \rangle \;=\;
(-i)^{2n+1}\left[\frac{\delta^n {\cal W}}{\delta J(x_1)\cdots \delta J(x_{2n})}\right]_{J=0}
\;.
\ee
The basic idea in forming connected functions is that each variation of a source pulls down an external graviton,
while the Taylor expansion of the Lagrangian provides for the rest of the structure, the vertices and propagators.

Now consider that any of the connected functions of interest must contain exactly two derivatives and, therefore, must
be linear in the Ricci scalar. Hence, each such amplitude contains exactly {one} factor of $a \sqrt{-g}{\cal R}$. But this is all that it can contain because, in the high-momentum regime, the cosmological constant is of no consequence. The only opportunity for the constant to enter into the formalism  is through an internal  propagator but, as discussed in $\S$~\ref{step2},
its presence never impacts upon  the calculations.
And so the connected $2n$-point function can have only  a single vertex, which is given by $a\sqrt{-g}{\cal R}$ expanded to
order $2n$ in number of gravitons. But, as discussed in Section~3, this is precisely what defines the 1PI $2n$-point  function!
To summarize,
\bea
\langle h^{2n} \rangle^E_{Con} &=& a\left[\sqrt{-g} {\cal R}\right]_{{\cal O}[h^{2n}]}
\nonumber \\
&=&
\langle h^{2n} \rangle^E_{1PI} \;.
\eea

Verifying this identity for all $n$ would mean showing that every conceivable 1PR function vanishes, a futile task!
Nonetheless,  one such example is presented
in Appendix~A.

A similar reasoning  applies to Gauss--Bonnet's order-$\epsilon$  corrections. In this case, the
Lagrangian is\; $a{\cal R}+b\Lambda+\epsilon {\cal R}_{\mu\nu\gamma\rho}{\cal R}^{\mu\nu\gamma\rho}$\;,
where we have used a field redefinition \cite{prop1,prop2,cavepeeps} to  transform the Gauss--Bonnet combination. Now, since these connected diagrams must contain exactly four derivatives and one power of $\epsilon$,
they are necessarily linear in  $\epsilon\sqrt{-g}{\cal R}_{\mu\nu\gamma\rho}{\cal R}^{\mu\nu\gamma\rho}$.
There is no room to include two derivatives from a Ricci scalar and, as before,
no opportunity to incorporate the cosmological constant. In equation,
\bea
\langle h^{2n} \rangle^{GB}_{Con} &=&
\epsilon\left[\sqrt{-g}{\cal R}_{\mu\nu\gamma\rho}{\cal R}^{\mu\nu\gamma\rho}\right]_{{\cal O}[h^{2n}]}
\nonumber \\
&=&
\langle h^{2n} \rangle^{GB}_{1PI}\;.
\eea

However, let us emphasize that limiting to linear order in $\epsilon$
is critical to the last conclusion. Had we, for instance, extended to order $\epsilon^2$, then
a connected function could  be constructed out of a pair of  Riemann-tensor-squared terms.
This is already evident for the simplest case of an $\epsilon^2$-order 1PR six-point function. Here, we start with the basic form
\be
\langle 6 \rangle^{\epsilon^2}\;=\;
\lim_{r,\tilde{r} \to \infty}\int dr \int d^4x \int d\tilde{r}\int d^4 \tilde{x}
\sqrt{-g(r,{\bf x})}\sqrt{-g(\tilde{r},\tilde{{\bf x}})}\;\nabla_d h(r,{\bf x})
\nonumber \ee \be
\times\; \nabla^d h(r,{\bf x}) \;
 \nabla_e h(r,{\bf x}) \;
 \Big\langle
\nabla^e h(r,{\bf x})\;\widetilde{\nabla^{f}} h(\tilde{r},\tilde{{\bf x}}) \Big\rangle\;
\widetilde{\nabla_{f}}h(\tilde{r},\tilde{{\bf x}})
 \;\widetilde{\nabla_{c}}h(\tilde{r},\tilde{{\bf x}})\;\widetilde{\nabla^{c}}h(\tilde{r},\tilde{{\bf x}}) \;
\;,
\ee
then integrate by parts
\be
\langle 6 \rangle^{\epsilon^2}\;=\;
-2\lim_{r,\tilde{r} \to \infty}\int dr \int d^4x \int d\tilde{r}\int d^4 \tilde{x}
\sqrt{-g(r,{\bf x})}\sqrt{-g(\tilde{r},\tilde{{\bf x}})}\;\nabla_d h(r,{\bf x})
\nonumber \ee \be
\times\; \nabla^d h(r,{\bf x})\; \nabla_e h(r,{\bf x})\;
 \Big\langle
\nabla^e h(r,{\bf x}) \;h(\tilde{r},\tilde{{\bf x}}) \Big\rangle\;
\widetilde{\nabla_{f}}h(\tilde{r},\tilde{{\bf x}})
 \;\widetilde{\nabla_{c}}h(\tilde{r},\tilde{{\bf x}})\;\widetilde{\nabla^{f}}\;\widetilde{\nabla^{c}}h(\tilde{r},\tilde{{\bf x}}) \;,
\ee
apply the inverse product rule and again  integrate by parts
\be
\langle 6 \rangle^{\epsilon^2}\;=\;
\frac{2}{3}\lim_{r,\tilde{r} \to \infty}\int dr \int d^4x \int d\tilde{r}\int d^4 \tilde{x}
\sqrt{-g(r,{\bf x})}\sqrt{-g(\tilde{r},\tilde{{\bf x}})}\;\nabla_d h(r,{\bf x})
\nonumber \ee \be
\times\; \nabla^d h(r,{\bf x}) \;
 h(r,{\bf x})\;
 \nabla_e\nabla^e\Big\langle
h(r,{\bf x}) \;h(\tilde{r},\tilde{{\bf x}}) \Big\rangle\;
\widetilde{\nabla_{f}}h(\tilde{r},\tilde{{\bf x}})
 \;\widetilde{\nabla_{c}}h(\tilde{r},\tilde{{\bf x}})\;\widetilde{\nabla^{f}}\;\widetilde{\nabla^{c}}h(\tilde{r},\tilde{{\bf x}}) \;,
\ee
employ  Eq.~(\ref{green}) to  integrate over the $(\tilde{r},\tilde{{\bf x}})$ coordinates
\be
\langle 6 \rangle^{\epsilon^2}\;=\;
\frac{2}{3}\lim_{r\to \infty}\int dr \int d^4x
\sqrt{-g(r,{\bf x})}\;\nabla_d h(r,{\bf x}) \;\nabla^d h(r,{\bf x}) \;h(r,{\bf x})
\nonumber \ee \be \times\;
\nabla_{f} h(r,{\bf x})
\;\nabla_{c}h(r,{\bf x})\;\nabla^f\;\nabla^c h(r,{\bf x}) \;,
\ee
and, finally, use integration  by parts followed by  the inverse product rule
to arrive at
\be
\langle 6 \rangle^{\epsilon^2}\;=\;
-\frac{2}{15}\lim_{r\to \infty}\int dr \int d^4x
\sqrt{-g(r,{\bf x})}\;\nabla_d h(r,{\bf x}) \;\nabla^d h(r,{\bf x})\;\nabla^f h(r,{\bf x}) \nonumber \ee \be \times \;
\nabla_{f} h(r,{\bf x})
\;\nabla_{c}h(r,{\bf x})\;\nabla^c h(r,{\bf x}) \;.
\ee

The Feynman relative weight  and Step~2 bring in another factor of $\;10/4=5/2\;$,
but the essential point is that this is only possible arrangement of derivatives ---
 there can be no cancellation! Meanwhile, the corresponding 1PI function is provided by
the next order in the Lovelock expansion, the six-derivative correction to Einstein gravity.

The general rule is that, at order $\epsilon^k$, the 1PI function is  given by the $2(k+1)$-derivative Lovelock
extension, whereas 1PR functions can be constructed out of any number of lower-order Lovelock terms (including the Einstein-theory term) provided that the total number of derivatives adds up to the same amount of $2(k+1)$ (anything less would be in violation
of the high-momentum regime). All this is a consequence of Einstein gravity being the sole two-derivative
theory of gravity and, thus,  the only one with a built-in mechanism to prevent against the proliferation
of derivatives in its connected amplitudes.

Finally, let us readdress the issue of neglecting vector and scalar modes in the internal lines. Internal vector modes are inconsequential by  the  same reasoning that we applied to the internal tensors. This is because the trace term in the propagator  is just as
 irrelevant for the vectors as it is for the tensors, from which
the rest of the argument follows in parallel.

But having scalar modes on an internal line is
a different matter. For scalars, the trace term in the propagator is now relevant to Step~2 and thus allows for the cosmological constant to contribute to amplitudes. Nonetheless, we maintain
that scalar modes are unphysical  in the absence of any external source.

To understand this, let us consider the simple case
of a pair of
three-point functions convolved into a four-point amplitude such that both internal gravitons are scalars. Up to an overall constant, the Einstein three-point function with a single scalar and two tensors goes as (with $H$ used to indicate the scalar mode)
\bea
\langle 3 \rangle_{hhH} \; = \;
\lim_{r \to \infty}\int dr \int d^4x
\sqrt{-g}\Big[&-& \frac{1}{2}\nabla^e h  \; \nabla_e h \; H^a_{\ a}\;+\;
\nabla^e h \;\nabla_f h \; H^f_{\ e}\Big. \nonumber \\
\Big. &+& h  \; h \; \nabla^e \nabla_e H^a_{\ a}\;-\;
 h \; h \; \nabla^e\nabla_f   H^f_{\ e}\;\;\Big]\;. \nonumber \\ & &
\eea

We next make the gauge choice $\;\nabla^a H^b_{\ a}= \alpha\nabla^b H^a_{\ a}\;$, where $\alpha$ will eventually be set to $\alpha=1$. This gauge choice is consistent with the previous choice of radial gauge. Then, in similar fashion to
previous calculations, we can use integration by parts and the inverse of the product rule to manipulate the above into
a single term. For instance,
\be
\langle 3 \rangle_{hhH} \; = \;
\lim_{r \to \infty}\frac{3}{4}(1-\alpha)\int dr \int d^4x
\sqrt{-g} \; h  \; h \; \nabla^e \nabla_e H^a_{\ a}\;.
\ee

For $\;\alpha=1\;$, the three-point function  vanishes.
So, $\langle 3 \rangle_{hhH}$ is unphysical and, therefore, cannot play a role in the calculation  of a physical
quantity such as the connected four-point function.
We could have, just as well,  first computed the 1PR contribution
to  the connected four-point and then imposed the gauge; the same outcome and conclusion would persist. For higher-point functions,
the procedure is expected to be more complicated but, without a source,
the outcome should be similarly harmless.

Our  intention is to use the results of this paper in
an AdS/CFT context and, therefore, a string-theory framework. Then we do need  to  consider the situation when a source for the scalar gravitons is present. More generally, we should consider a possible source for the scalars that
originates from compactifications of string theory. In this event,  such contributions can no longer be gauged away and, as already mentioned, the cosmological constant term could be relevant to the computation of the connected amplitudes. What saves us from this enormous complication is that the additional contributions are higher order in $\epsilon$.

In the effective field theory of gravity that is induced by string theory,
 any  such source for scalars comes with a coupling that scales at least as order $\epsilon$ \cite{morerob}. Hence, by introducing  a scalar onto an internal line,  one
is suppressing the  amplitude by  another factor of
$\epsilon$ without increasing the number of derivatives.
Hence, for the high-momentum region in particular,  a physical scalar mode is  subleading to the connected multi-point functions of the tensor gravitons.

\section{Disconnected functions}

Although neglected so far,  disconnected graviton functions
can provide a useful check on consistency
when it comes time to match our theoretical predictions with experiment.

To see how this works, let us first consider the disconnected portion of
the  four-point function
\be
\langle hhhh\rangle_{Dis} \;=\;\frac{4!}{(2!)^2}\langle hh \rangle^2\;,
\label{top}
\ee
where the ratio of factorials  accounts for the differing symmetrization factors. This is shown in Fig.~8a.

As the above relation makes clear, one can use the value of the disconnected four-point function as an alternative means for evaluating the propagator. All that is required is the relevant combinatorics, so as to distinguish between the  disconnected and connected portions of the full four-point function.
As per the discussion in Subsection~\ref{combo}, the ``Feynman weight'' of the disconnected part is determined by the number of ways of contracting   $h^2\cdot h^4 \cdot h^2$ such that connectivity is broken but without any loops. The answer is $\;4!/2!\;$,~\footnote{To avoid loops, the $h^2$'s can only contract with the external $h^4$, leading to $4!$\;.
The other factor of $1/2!$ comes from the expansion of the exponent.}
and so a relative weight (in comparison to the
connected (equivalently, 1PI) four-point function of
$\;\frac{4!}{2!}\frac{1}{4!}=\frac{1}{2!}\;$.  Hence,
\be
\frac{4!}{(2!)^3}\langle hh \rangle^2 \;=\; \langle hhhh \rangle_{Ful}-\langle hhhh\rangle_{Con}
\ee
or
\be
\langle hh \rangle \;=\; \frac{1}{3}\frac{\left[\langle hhhh \rangle_{Ful}-\langle hhhh\rangle_{Con}\right]}{\langle hh \rangle}\;,
\ee
where the denominator on the right-hand side side now indicates
{division} by a two-point function and { not} an inverse propagator.

\begin{figure}[t]
\vspace{-0.5in}\hspace{-.5in}
\scalebox{.55}{\includegraphics[angle=-90,trim=0 0 0 0]{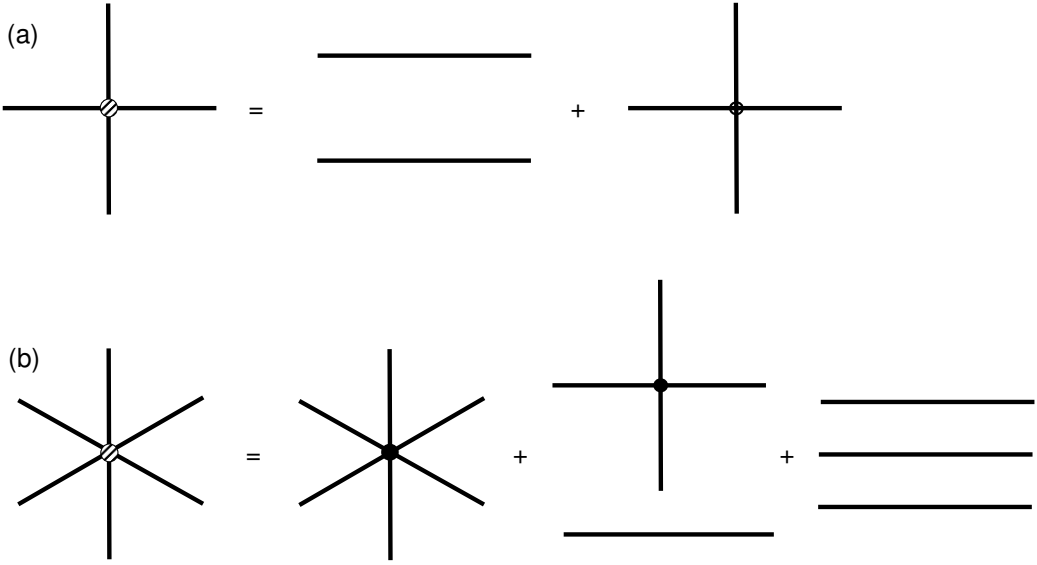}}
\vspace{-.9in}
\caption{Full functions as sums of their
connected and disconnected components.}
\end{figure}

Also having utility is the disconnected portion of the six-point function,
\be
\langle hhhhhh\rangle_{Dis}\; =\;\frac{6!}{1!}\frac{6!}{4!\cdot 2!}\langle hhhh \rangle_{Con}
\langle hh \rangle\;\;+\;\frac{6!}{3!}\frac{6!}{(2!)^3}\langle hh \rangle^3 \;,
\ee
where the  first fraction in front of either  term represents the Feynman
weight (as per the previous paragraph)
and the second fraction accounts for the symmetrization factors
(as per Eq.~(\ref{top})). These are depicted in Fig.~8b.

Now, suppose that the intent is to find
an alternative  means of deducing the connected four-point function.
Then we should be looking at
\be
\frac{1}{1!}\frac{6!}{4!\cdot2!}\langle hhhh\rangle_{Con}\langle hh\rangle\;=\;
\langle hhhhhh \rangle_{Ful}
-\langle hhhhhh\rangle_{Con}-\frac{1}{3!}\frac{6!}{(2!)^3}\langle hh \rangle^3
\ee
or
\be
\langle hhhh\rangle_{Con}\;=\; \frac{1}{15}\frac{\left[\langle hhhhhh
\rangle_{Ful}
-\langle hhhhhh\rangle_{Con}\right]}{\langle hh\rangle}-\langle hh\rangle^2\;.
\ee

\section{Conclusion}

To summarize, we have shown how to
translate our previous results on the
1PI graviton amplitudes \cite{cavepeeps}
into their connected-function counterparts. This is an
important ingredient if the gauge--gravity duality is to be used
for determining the corresponding  stress-tensor correlations functions on the field-theory side. As explained in \cite{newby}, such correlation functions would then provide the means for experimentally probing the gravitational dual  of a strongly coupled fluid.

We have shown by explicit calculation that a large class of the 1PR connected diagrams cancel off and have argued that this cancelation persists for all 1PR contributions in the high-momentum regime for both Einstein gravity and
its leading-order correction.

In many instances, a vanishing outcome where it was not expected can be
the result of a symmetry principle. Is there such a principle here?
We are looking at a  specific kinematic region for which the radial derivatives are neglected and in essence are assumed to vanish. Away from the high-momentum regime,  the radial derivatives will have some non-vanishing value
which would correspond to spontaneously breaking the unknown associated symmetry. In this sense, the radial degree of freedom could be viewed as a Goldstone mode.

From the perspective of the boundary gauge  theory, the radial coordinate represents an energy scale. Radial differentiation corresponds to a flow in this scale.  That the flow has been rendered  inert  suggests a conformal fixed point of the gauge theory. Hence, the observed  cancelations could indicate an unbroken conformal symmetry of the boundary theory for a finite $N$.

\section*{Acknowledgments}

The research of RB was supported by the Israel Science Foundation grant no. 239/10. The research of AJMM  was  supported by  a  Rhodes University Discretionary Grant RD11/2012. AJMM thanks Ben Gurion University for their hospitality during his visit.

\appendix

\section{``Four-Four-Four''}

Here, it will be shown that, at least in one instance,
a reducible function consisting of more than two components ({\it i.e.},  more than one internal line)
does indeed vanish. Out test case is the convolution of three four-point functions
into an eight-point function,
\be
\langle h^8 \rangle_{ {4-4-4}} \;=\;
\frac{\langle hhhh\rangle_{1PI} \langle hhhh \rangle_{1PI} \langle hhhh \rangle_{1PI}}{\langle hh
\rangle \langle hh \rangle}\;.
\ee

We consider Einstein's theory and start by convolving the interior four-point function with one on the exterior.  This is a different calculation than that of $\S$~\ref{einsix} because the interior
four-point function has { only two external gravitons}.
Hence, this is more akin to a $2p$-$2q$ ($p\neq q$) convolution.
Also, it is now important to keep careful track of the various structures.

This first convolution consists of  eight distinct cases, which then leads to sixteen different diagrams depending on how the derivatives are initially arranged.
We will work through one case in detail and then report the findings of the other seven.

\begin{figure}[t]
\vspace{-1.5in}\hspace{+0.06cm}
\scalebox{.45}{\includegraphics[angle=-90]{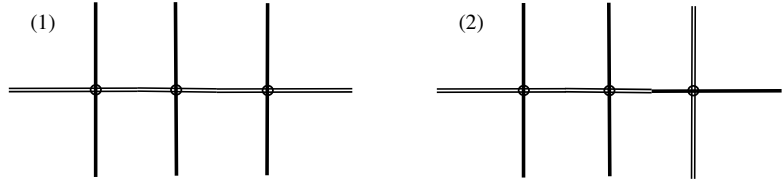}}
\hspace{-.5in}
\caption{Two diagrams of 4-4-4 depicting the case discussed in detail in the text. The initial convolution is between the left-most and interior four-point functions.
 The difference between the two diagrams is in the right-most part of the diagram where the double lines and single lines are exchanged. }
\end{figure}

Let us thus consider the two diagrams in Fig.~9,
\be
\label{eq88}
\langle 6^{\ast} \rangle^{(1,2)} \;=\;
\lim_{r,\tilde{r} \to \infty}\int dr \int d^4x \int d\tilde{r}\int d^4 \tilde{x}
\sqrt{-g(r,{\bf x})}\sqrt{-g(\tilde{r},\tilde{{\bf x}})}\;h(r,{\bf x})
\nonumber \ee \be
\times\;h(r,{\bf x})\; \nabla_e h(r,{\bf x}) \;
 \Big\langle
\nabla^e h(r,{\bf x})\;\widetilde{\nabla^{f}} h(\tilde{r},\tilde{{\bf x}}) \Big\rangle\;
\widetilde{\nabla_{f}} h^{\ast}(\tilde{r},\tilde{{\bf x}})
 \;h(\tilde{r},\tilde{{\bf x}})\;h(\tilde{r},\tilde{{\bf x}}) \;,
\ee
where the superscript on the left side denotes the particular diagram(s) and the meaning of asterisk on the graviton is the following: The interior four-point function contains two internal gravitons, one of which is to be contracted in the convolution of Eq.~(\ref{eq88}) and one of which remains  to be contracted in the convolution to follow. The former is, as usual, included within the expectation value, while the latter is marked with an asterisk for future reference.

Starting with integration on the left (the $({r},{{\bf x}})$ coordinates) and applying the usual manipulations, we arrive at  two distinct terms, denoted as $b_1$ and $b_2$, corresponding respectively to diagrams~(1) and~(2) in Fig.~9:
\be
b_1\;=\;\beta_1\lim_{\tilde{r} \to \infty} \int d\tilde{r}
\int d^4 \tilde{x}
\sqrt{-g(\tilde{r},\tilde{{\bf x}})}\;h(\tilde{r},\tilde{{\bf x}})\;h(\tilde{r},\tilde{{\bf x}})
\;h_S(\tilde{r},\tilde{{\bf x}})
\nonumber \ee \be
\times\;
\widetilde{\nabla_{f}}\; h^{\ast}_S(\tilde{r},\tilde{{\bf x}})
\;\widetilde{\nabla^{f}} h(\tilde{r},\tilde{{\bf x}})
 \;h(\tilde{r},\tilde{{\bf x}})
\label{dee}\;,
\ee
\be
b_2=\;\beta_2\lim_{\tilde{r} \to \infty} \int d\tilde{r}\int d^4\tilde{x}
\sqrt{-g(\tilde{r},\tilde{{\bf x}})}\;h(\tilde{r},\tilde{{\bf x}})\;h(\tilde{r},\tilde{{\bf x}})
\;h_S(\tilde{r},\tilde{{\bf x}})
\nonumber \ee \be
\times\;
\widetilde{\nabla^{f}} \;\widetilde{\nabla_{f}} h^{\ast}_S(\tilde{r},\tilde{{\bf x}})
\;h(\tilde{r},\tilde{{\bf x}})
 \;h(\tilde{r},\tilde{{\bf x}})\;, \label{dum}
\ee
where  $\beta_1$ and $\beta_2$ are relative weights which are determined below.

Notice that the process of tensor contraction has, for either $b_1$ or $b_2$, induced a symmetric contraction between the surviving internal graviton and its new partner. (The end product of Step~2 is
always a symmetric contraction; {\em cf}, $\S$~\ref{step2}.)  Because of the various integrations by parts, the distinction between symmetric and anti-symmetric contractions is no longer as simple as looking for  which gravitons are differentiated.  As far as the surviving internal graviton is concerned, this distinction  is  important in the sequel, and  so  we label both it and its new partner with a subscript of $S$ or $A$ accordingly.

Next,  $b_1$ and $b_2$ are assigned the  respective relative weights of
\be
\beta_1\;=\;\frac{3}{5}\cdot 2\cdot\frac{1}{3}\cdot \frac{1}{4}\cdot \frac{1}{2} \cdot\frac{1}{6} \;=\;\frac{1}{120}\;,
\ee
\be
\beta_2\;=\;\frac{3}{5}\cdot 1 \cdot \frac{1}{3}\cdot \frac{1}{4}\cdot \frac{1}{2} \cdot\frac{1}{6} \;=\;\frac{1}{240}\;.
\ee

Let us explain how these weights are obtained,  focusing on $\beta_1$ ($\beta_2$ follows similarly). From left to right, $\frac{3}{5}$ is
due to our symmetrization rule, the next two numbers are a consequence of integrating by parts (2 from the product rule and
$\frac{1}{3}$ from  the inverse of the  product rule),
$\frac{1}{4}$ is the contribution from Step~2, $\frac{1}{2}$ is the relative ratio for the case in which the left-most internal graviton is differentiated and $\frac{1}{6}$ is the relative ratio for the case
in which both of  the right-side internal  gravitons are differentiated.

If we, rather, integrate over the $(\tilde{r},\tilde{{\bf x}})$ coordinates first, there is only one result,
\be
b_3\;=\;\beta_3\lim_{r \to \infty} \int dr\int d^4x \sqrt{-g(r),{\bf x})}\;h(r,{\bf x})\;\nabla^e h(r,{\bf x})
\;\nabla_e h_S(r,{\bf x})
\nonumber \ee \be
\times\; \
h^{\ast}_S (r,{\bf x})
\;h(r,{\bf x})\;h(r,{\bf x})
\;,
\ee
with a weight of
\be
\beta_3\;=\;\frac{2}{5}\cdot 2\cdot \frac{1}{3}\cdot \frac{1}{4}\cdot \frac{1}{2} \cdot\frac{1}{6} \;=\;\frac{1}{180}\;.
\ee

\begin{figure}[t]
\vspace{-0.5in}\hspace{-8.6cm}
\scalebox{.45}{\includegraphics[angle=-90]{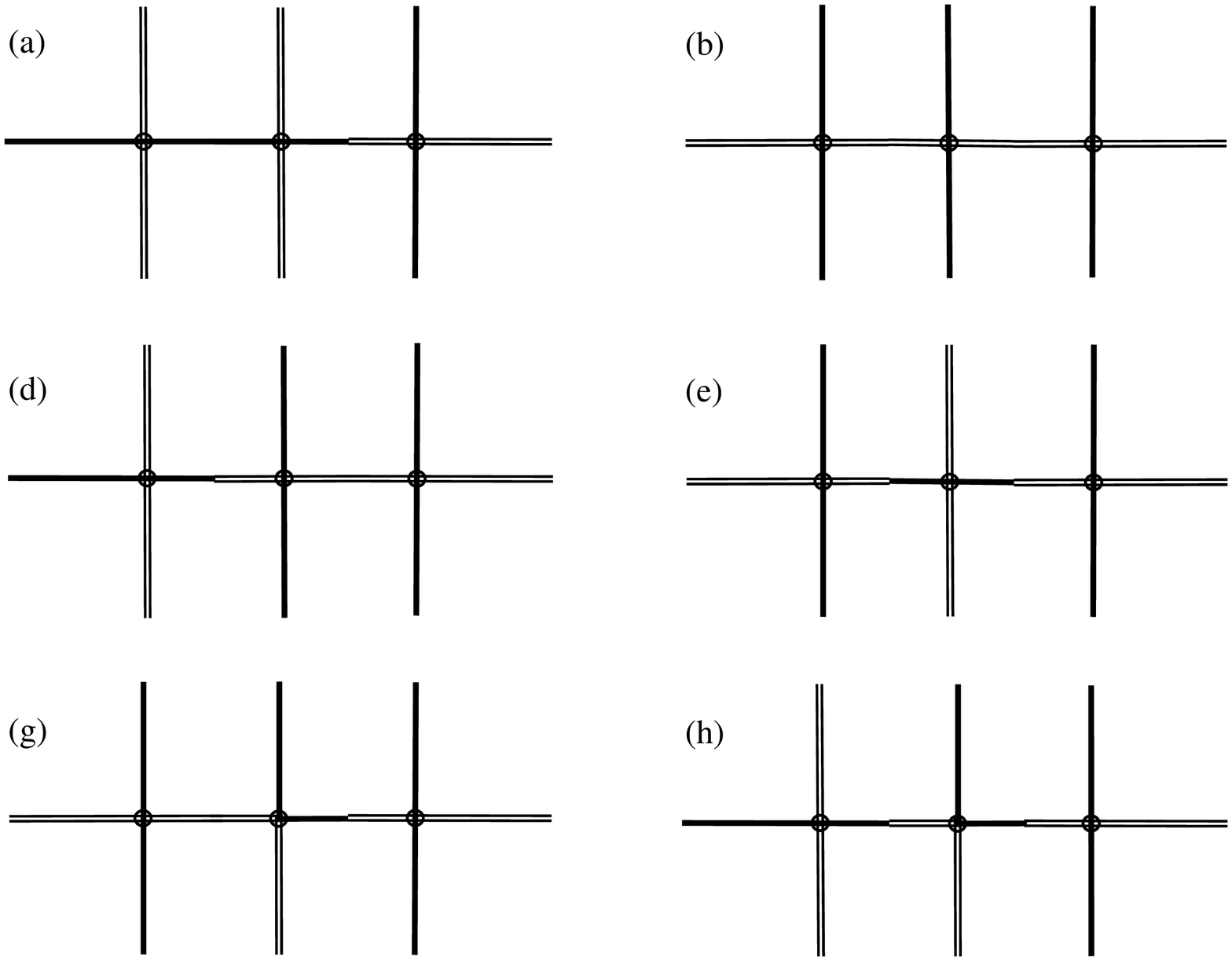}}
\hspace{-0.5in}
\scalebox{.45}{\includegraphics[angle=-90,trim=0 0 0 1000]{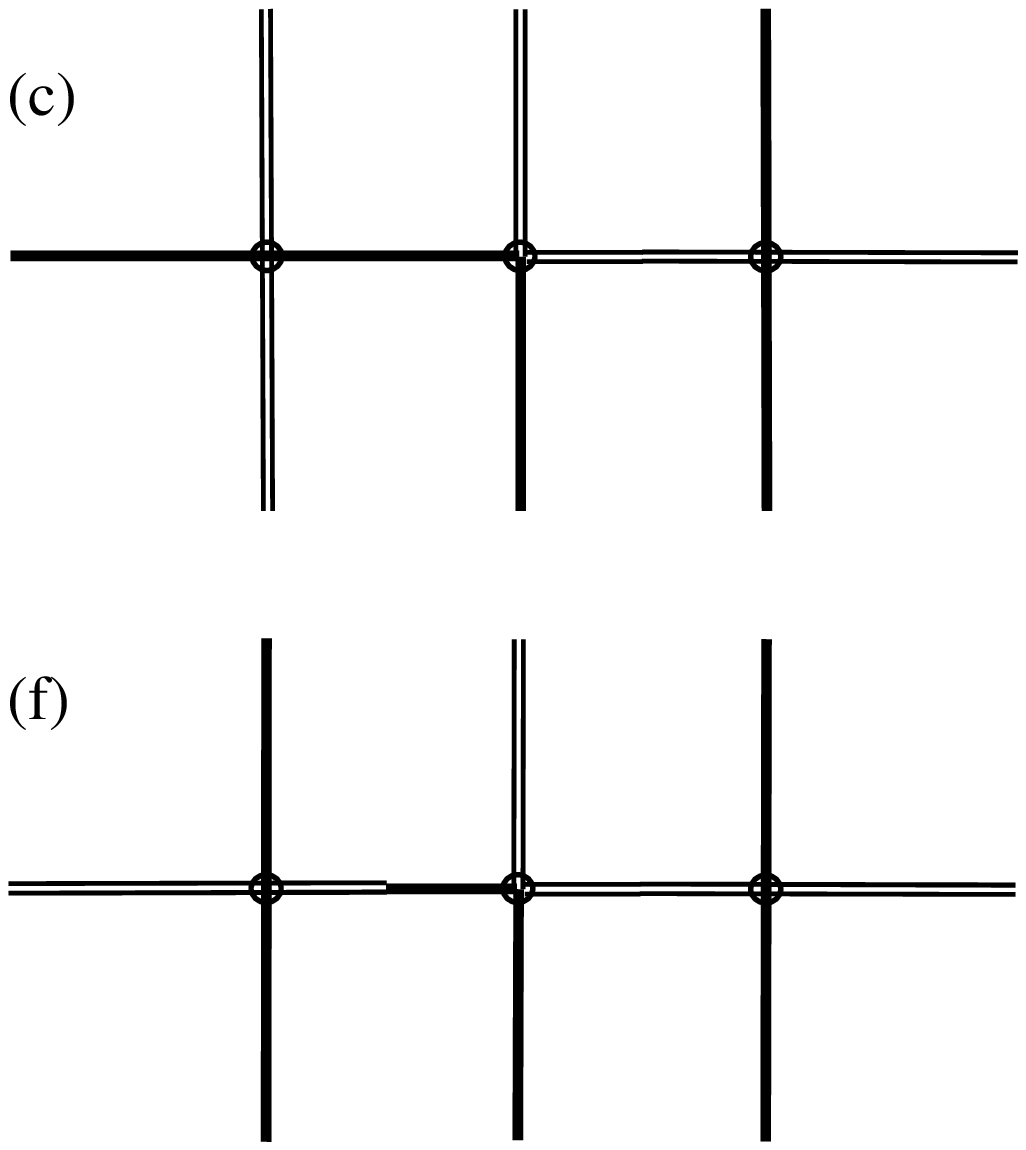}}
\caption{Eight diagrams of 4-4-4. Each of the diagrams represents one of the cases in the text, with the initial convolution being between the left-most and interior four-point functions. The right-most four-point function depicts the last term in Eq.~(\ref{rightext}). First row, from left to right, cases (a-c), second row, cases (d-f), third row, cases (g-h). Each of the diagrams has a partner diagram (not shown), where the double lines and single lines on the right are exchanged as in Fig.~9.}
\end{figure}

Let us summarize this  case using abridged notation and
 call it case~(b) as represented in Fig.~10: \\
{\bf case b}
\bea
\langle 6^{\ast}\rangle^{(b)} & = &
\int\int \;h \;h \;\nabla h \;\Big\langle \nabla h \;\nabla h \Big\rangle \;\nabla  h^{\ast}\; h \;h
\nonumber \\
& =&
\frac{1}{120}\int \;h \;h \;h_S \;\nabla  h^{\ast}_S \;\nabla h \; h
\;+\frac{1}{240}\int \;h  \;h  \;h_S \;\nabla \nabla  h^{\ast}_S \;  h  \;h
\nonumber \eea \be
\;+\;\frac{1}{180}\int \;h  \;\nabla h \;\nabla h_S \;  h^{\ast}_S  \;h  \;h\;.
\ee

The other seven cases and their outcomes go as follows: \\
{\bf case a}
\bea
\langle 6^{\ast} \rangle^{(a)} & = &
\int\int \;\nabla h \;\nabla h \; h \;\Big\langle h  \;h  \Big\rangle  \;h^{\ast} \;\nabla h \;\nabla h
 \nonumber \\ & = &
\frac{1}{72}\int \;\nabla h \;\nabla h \; h_S \;h^{\ast}_S \;h  \;h  \;,
\eea
{\bf case c}
\bea
\langle 6^{\ast} \rangle^{(c)} & = &
\int\int \;\nabla h \;\nabla h \; h \;\Big\langle h  \;h  \Big\rangle  \;h \;\nabla h \;\nabla h^{\ast}
\nonumber  \\ & = &
\frac{1}{60}\int  \;h   \;h  \;h \;h  \;\nabla h_A \;\nabla  h^{\ast}_A
\;+\;\frac{1}{90}\int \;\nabla h \;\nabla h \; h \;h  \;h_A  \;h^{\ast}_A  \;,
\nonumber \\ &  &
\eea
{\bf case d}
\bea
\langle 6^{\ast} \rangle^{(d)} & = &
\int\int \;\nabla h \;\nabla h \; h \;\Big\langle h \;\nabla h  \Big\rangle \;\nabla h^{\ast}
\;  h \;h
 \nonumber \\ & = &
-\frac{1}{120}\int \;h \;h \;h_S \;\nabla  h^{\ast}_S\; h \;\nabla h
\;-\;\frac{1}{240}\int \;h \;h \;h_S \;\nabla\nabla h^{\ast}_S \; h \;h
\nonumber \eea \be
\;-\;\frac{1}{180}\int  \;\nabla h \;\nabla h  \; h_S \;h^{\ast}_S \;h \;h \;,
\ee
{\bf case e}
\bea
\langle 6^{\ast} \rangle^{(e)} & = &
\int\int \;h \;h \;\nabla h \;\Big\langle \nabla h \; h \Big\rangle\; h^{\ast} \;\nabla h \;\nabla h
\nonumber \\ & = &
-\frac{1}{120}\int \;h \;h \;h_S \;h^{\ast}_S \;\nabla h \;\nabla h
\;-\;\frac{1}{180}\int \;h \;\nabla h \;\nabla h_S \;  h^{\ast}_S  \;h  \;h\;,
\nonumber \\ & &
\eea
{\bf case f}
\bea
\langle 6^{\ast} \rangle^{(f)} & = &
\int\int \;h \;h \;\nabla h \;\Big\langle \nabla h \; h \Big\rangle\; h \;\nabla h \;\nabla h^{\ast}
 \nonumber \\ & = &
-\frac{1}{60}\int \;h \;h \;h \;h \;\nabla h_A\;\nabla h_A^{\ast}
\;-\;\frac{1}{90}\int \;\nabla h \;\nabla h \;h \;h_A \;h^{\ast}_A \;,
\nonumber \\ & &
\eea
{\bf case g}
\bea
\langle 6^{\ast} \rangle^{(g)} & = &
\int\int \;h \;h \;\nabla h \;\Big\langle \nabla h \;\nabla h \Big\rangle \;\nabla h \; h \;h^{\ast}
\nonumber\\ & = &
\frac{1}{120}\int \;h \;h \;h \;\nabla h \;\nabla h_S \; h^{\ast}_S
\;+\;\frac{1}{120}\int \;h \;h \;h \;\nabla h \; h_S \;\nabla  h^{\ast}_S
\nonumber \eea \be
\;+\;\frac{1}{90}\int \;h \;\nabla h \;\nabla h \; h \;h_S \;h^{\ast}_S \;,
\ee
{\bf case h}
\bea
\langle 6^{\ast} \rangle^{(h)} & = &
\int\int \;\nabla h \;\nabla h \; h \;\Big\langle h \;\nabla h  \Big\rangle \;\nabla  h\;h \;h^{\ast}
\nonumber \\ & =  &
-\frac{1}{90}\int \;\nabla h \;\nabla  h \; h \;h \;h_S \;h^{\ast}_S
\;-\;\frac{1}{60}\int \;h \;h \;\nabla h \; h \;h_S  \;\nabla h^{\ast}_S  \;.
\nonumber \\ &  &
\eea

Summing up the results of the eight cases, we find just a pair of terms,
\bea
\langle 6^{\ast} \rangle  \; =  \;
\;\frac{1}{120}\int \;h \;h \;h \;\nabla h \;\nabla h_S \; h^{\ast}_S
\;-\frac{1}{120}\int \;h\;h\;h \;\nabla h \; h_S \;\nabla h^{\ast}_S  \;.
\nonumber \\
\eea

It is straightforward to show that the convolution of either of these with the remaining four-point function,
\be
\langle 4^{\ast} \rangle\;=\; \frac{1}{2}\int \;\nabla h \;\nabla h\; h_S \;h^{\ast}_S
\;+\frac{1}{2}\int \;h \;h  \;\nabla h_A \;\nabla h^{\ast}_A \;,
\label{rightext}
\ee
is vanishing. These calculations  follow exactly as in  $\S$~\ref{pq} with two caveats. First, for the purposes of Step~2, if an internal graviton is labeled with an $S$ ($A$), it is treated as undifferentiated (differentiated).  Second, relative-ratio factors have  already been assigned. This is also true for the four-point function when expressed in the above form.

Finally, since $\;\langle h^8 \rangle_{ {6-4}}=0\;$ from the considerations of
Section~6,  the preceding result allows us to conclude
that $\;\langle h^8 \rangle_{Con}=\langle h^8 \rangle_{1PI}\;$ at least to order $\epsilon^0$.

\end{document}